\documentclass[twocolumn,aps,superscriptaddress,showpacs,floatfix]{revtex4}
\usepackage{graphicx}
\usepackage{dcolumn}
\usepackage{bm}
\usepackage[colorlinks,linkcolor=blue,anchorcolor=blue,citecolor=blue,urlcolor=blue]{hyperref}
\usepackage{amsmath}
\usepackage{multirow}
\usepackage{booktabs}  
\usepackage{longtable}  
\usepackage{float}
\usepackage{array}
\usepackage{etoolbox}
\usepackage{makecell,rotating,multirow,diagbox}
\usepackage{setspace}
\usepackage{tabularx}
\usepackage[draft]{changes}
\usepackage[section]{placeins}
\usepackage{lineno,hyperref}
\usepackage{supertabular}
\usepackage{amssymb}
\usepackage[normalem]{ulem}
\usepackage[colorlinks,linkcolor=blue,anchorcolor=blue,citecolor=blue,urlcolor=blue]{hyperref}
\renewcommand\sout{\bgroup \color{red} \ULdepth=-.5ex \ULset}

\begin{document}

\title{Systematic study of cluster radioactivity in trans-lead nuclei within various versions of proximity potential formalisms}

\author{Xiao Liu}
\affiliation{School of Nuclear Science and Technology, University of South China, 421001 Hengyang, People's Republic of China}
\author{Jie-Dong Jiang}
\affiliation{School of Nuclear Science and Technology, University of South China, 421001 Hengyang, People's Republic of China}
\author{Xi-Jun Wu}
\email{wuxijunusc@163.com }
\affiliation{School of Math and Physics, University of South China, 421001 Hengyang, People's Republic of China}
\author{Xiao-Hua Li}
\email{lixiaohuaphysics@126.com }
\affiliation{School of Nuclear Science and Technology, University of South China, 421001 Hengyang, People's Republic of China}
\affiliation{Cooperative Innovation Center for Nuclear Fuel Cycle Technology $\&$ Equipment, University of South China, 421001 Hengyang, People's Republic of China}
\affiliation{National Exemplary Base for International Sci $\&$ Tech. Collaboration of Nuclear Energy and Nuclear Safety, University of South China, Hengyang 421001, People's Republic of China}

\begin{abstract}
	
In this work, based on the framework of the Coulomb and proximity potential model (CPPM), we systematically study the cluster radioactivity half-lives of 26 trans-lead nuclei by considering the cluster preformation probability which is found to possess a simple mass dependence on the emitted cluster by R. Blendowske and H. Walliser [Phys. Rev. Lett. 61, 1930(1988)]. Meanwhile, we investigate 28 different versions of the proximity potential formalisms, which are the most complete known proximity potential formalisms and have been proposed for the description of proton radioactivity, two-proton radioactivity, $\alpha$ decay, heavy-ion radioactivity, quasi-elastic scattering, fusion reactions and other applications. 
The calculated results show that the modified forms of proximity potential 1977 denoted as Prox.77-12 and the proximity potential 1981 denoted as Prox.81 are the most appropriate proximity potential formalisms for the study of cluster radioactivity as the root-mean-square deviation between experimental data and relevant theoretical results obtained are least and the both values are 0.681. For comparison, a universal decay law (UDL) proposed by Qi \textit{et al.} [Phys. Rev. C 80, 044326 (2009)], a unified formula of half-lives for $\alpha$ decay and cluster radioactivity proposed by Ni \textit{et al.} [Phys. Rev. C 78, 044310 (2008)] and a scaling law (SL) in cluster decay proposed by Horoi \textit{et al.} [J. Phys. G 30, 945 (2004)] are also used.
In addition, utilizing CPPM with Prox.77-12, Prox.77-1, Prox.77-2 and Prox.81, we predict the half-lives of 51 potential cluster radioactive candidates whose cluster radioactivity is energetically allowed or observed but not yet quantified in NUBASE2020. The predicted results are in the same order of magnitude with those obtained by using the compared semi-empirical and/or empirical formulae. At the same time, the competition between $\alpha$ decay and cluster radioactivity of these predicted nuclei is discussed and it is found that $\alpha$ decay predominates by comparing half-lives.
 
\end{abstract}

\maketitle

\section{Introduction}
\setlength{\parskip}{0pt}
Spontaneous radioactivity of nuclei has always been an important and popular research field in nuclear physics, which firstly discovered by Becquerel in 1896 \cite{1903Becquerel}. Soon after that, Rutherford observed the spontaneous emission of $\alpha$ particles from the nuclei in the experiment and named as $\alpha$ decay \cite{1908PRSA81141, 1908PM1281}. It was not until 1928 that Gurdney and Condon, Gamow independently succeeded in providing a theoretical explanation of $\alpha$ decay using the tunneling effect of quantum mechanics \cite{1928ZP51204, 1928Nature122439, 1929PR33127}. Nowadays, spontaneous radioactivity of nuclei has been known to exist in many types \cite{2023NST3455, 2022NST33122, 2018PRC97044322, 2016PRC93034316, 2015ADNDT1011, 2023CPC47094103}. Cluster radioactivity is also one of them that occurs mainly in the regions of heavy nuclei, which is an intermediate process between $\alpha$ decay and spontaneous fission \cite{1988PRL611930, 2010NPA83838, 2010PRC82014607, 2009PRC80024310}. In this process, the parent nucleus emits a cluster particle that is heavier than an $\alpha$ particle but lighter than the lightest fission fragment, while decaying into a doubly magic daughter nucleus $^{208}$$\rm{Pb}$ or its neighboring daughter nucleus \cite{2021PS96125322, 2020PRC102034318, 1991JPG17S443, 1985PRC311984, 1998PR294265, 2010SPRINGER}.
This peculiar decay mode has aroused the interest and research of numerous physicists since it provides multitudinous vital information for studying nuclear structure \cite{1994IJMPE3335, 1987Nature325137, 1989ARNPS3919, 2007RRP592, 1989JPG15529}.
In 1980, S$\check{a}$ndulescu, Poenaru and Greiner made the first prediction of this radioactivity \cite{1980WGF111334}. In 1984, Rose and Jones experimentally observed the emission of $^{14}$$\rm{C}$ from $^{223}$$\rm{Ra}$, thus verifying this decay mode \cite{1984Nature307245}. Nowadays, an increasing number of clusters ranging from $^{14}$$\rm{C}$ to $^{34}$$\rm{Si}$ have been observed experimentally to be emitted from the parent nuclei ranging from $^{221}$$\rm{Fr}$ to $^{242}$$\rm{Cm}$ and their half-lives have been measured respectively \cite{1989PJP32419, 1985PRC322198, 1991PRC44888, 1992PRC461939}. 

So far, there are numerous models and/or approaches proposed to well comprehend cluster radioactivity, which are mainly divided into two extreme categories\cite{1984JPG10183, 1985PRC32572, 1986SJNP44923, 1985NPA438450, 1987NPA464205, 1988PRC381377, 2000P55375, 2012PRC86044612, 2004PRC70034304, 2013PRC87024308, 2015NPA93659, 2012PS86015201, 2017NPA958187, 2012PRC85044313, 2009PRC80037307, 2009EPJA41197, 2005PRC71014301}. 
The one is considered as a spontaneous fission process with super asymmetric mass in an adiabatic state, in which the parent nucleus continuously deforms until reaches the scission configuration after crossing the potential barrier \cite{1984JPG10183, 1985PRC32572, 1986SJNP44923, 1985NPA438450, 1987NPA464205, 1988PRC381377, 2000P55375}.
The other is considered as an $\alpha$-like process in a non-adiabatic state, in which the cluster particle is preformed in the parent nucleus with a certain probability and then penetrates the potential barrier \cite{2012PRC86044612, 2004PRC70034304, 2013PRC87024308, 2015NPA93659, 2012PS86015201, 2017NPA958187, 2012PRC85044313, 2009PRC80037307, 2009EPJA41197, 2005PRC71014301, 2023PRC108024306, 2024NPA1041122787}.
In the former, such as the Analytical Super Asymmetric Fission Model (ASAFM) of Poenaru \textit{et al.} \cite{1984JPG10183, 1985PRC32572}, the Cubic-plus-Yukawa-plus-Exponential Potential Model (CYEM) of Shanmugam and Kamalaharan \cite{1988PRC381377}, the Coulomb and proximity potential model (CPPM) of Santhosh \textit{et al.} \cite{2000P55375} and so on \cite{1986SJNP44923, 1985NPA438450, 1987NPA464205}, all can accurately reproduce the experimenttal data of the cluster radioactivity half-lives.
In the latter, it is well explained in the Preformed Cluster Model (PCM), where the cluster preformation probability is calculated through solving the Schr$\ddot{o}$dinger equation for the dynamic flow of charges and masses by Gupta and Malik\cite{2012PRC86044612}. Ren \textit{et al.} also provide a strong supporting proof for this type by considering the influence of charge number on the preformation factor under the framework of the microscopic densitydependent model (DDCM) with the renormalized M3Y nucleon-nucleon interaction to successfully calculate the cluster radioactivity half-lives \cite{2004PRC70034304} and so on \cite{2013PRC87024308, 2015NPA93659, 2012PS86015201, 2017NPA958187, 2012PRC85044313, 2009PRC80037307, 2009EPJA41197, 2005PRC71014301, 2023PRC108024306, 2024NPA1041122787}. At the same time, some valid empirical and/or semi-empirical formulae can be applied to calculate the half-life of cluster radioactivity, such as a universal decay law (UDL) proposed by Qi \textit{et al.} \cite{2009PRC80044326, 2009PRL103072501}, a unified formula of half-lives for $\alpha$ decay and cluster radioactivity proposed by Ni \textit{et al.} \cite{2008PRC78044310}, a scaling law (SL) in cluster decay proposed by Horoi \textit{et al.} \cite{2004JPG30945} and so on \cite{2013EPJA496, 2004PRC70017301, 2023CPC47064107, 2008JPG35085102, 2013EPJA4966, 2023EPJA59189, 2023NPA1031122597}.  

In 1970s, the proximity potential was firstly proposed by Blocki \textit{et al.} to deal with heavy-ion reactions \cite{1977AP105427}. It is a nucleus-nucleus interaction potential based on the proximity force theorem, which is described as the product of two parts \cite{1981AP13253}. The one is a factor determined by the mean curvature of the interaction surface, the other is a universal function that depends on the separation distance and is independent of the masses of colliding nuclei \cite{2010PRC81044615}. The concept of the universal function is fundamental advantage of proximity potential since it has the merits of simple and precise formalism. Nowdays, the proximity potential has developed variety versions with its own characteristics from the original version (Prox.1977) \cite{1977AP105427} by improving the surface energy coefficients, the universal function, nuclear radius parameterization and so on \cite{1967AF36343, 1966NP811, 1976NPA272502, 1979PRC20992, 1981NPA361117, 1984JPG101057, 1988ADNDT39213, 1995ADNDT59185, 2003PRC67044316, 2000PRC62044610, 2010CPL27112402, 2011P76921, 1973PLB47139, 1977PRL39265, 1994JPG201297, 1976PLB6519, 1995NPA594203, 1980NPA348140, 2002PLB526315, 2013NPA89754}, which have been applied to different fields for comparative study by nuclear physicists \cite{2015EPJA51122, 2022PS97095304, 2019EPJA5558, 2016PRC93024612, 2017EPJP132431, 2009NPA81735, 2014NPA922191, 2014JPG41105108, 2015PRC91044603, 2019NPA992121637}. For the field of cluster radioactivity, it has also been extensively researched \cite{2012PRC86044612, 2017EPJA53136, 2016NPA95186}. 
In 2012, Kumar \textit{et al.} conducted research showing that the proximity potential 1977 could be a better option for studying cluster radioactivity by using 8 different versions of proximity potential formalisms and the preformation probability obtained by solving the stationary Schr$\ddot{o}$dinger equation for the dynamic flow of mass and charge in preformed cluster model(PCM) \cite{2012PRC86044612}. In 2016, Zhang \textit{et al.} showed that the calculated results of Bass77 and Denisov potentials were most consistent with experimental data for large cluster radioactivity of even-even nuclei by comparing 14 different proximity potential formalisms \cite{2016NPA95186}. Soon after, Santhosh \textit{et al.} concluded that Bass80 was the most appropriate potential for studying cluster radioactivity by using a simple power-law interpolation to calculate the penetration probability inside the barrier as the preformation probability and comparing the calculations from 12 different versions of proximity potential formalisms \cite{2017EPJA53136}.
The most suitable proximity potential formalism for the study of cluster radioactivity, the above researchers obtained different conclusions, which may be due to discrepancies in the evaluation of cluster preformation probability, the different types and numbers of proximity potential formalisms adopted, the inclusion of nuclear deformation and so on when the cluster radioactivity half-lives are calculated. Therefore, it is necessary to explore the systematic behavior of more different proximity potential and to select the most promising proximity potential formalism for the cluster radioactivity. To this end, considering the preformation probability with a simple mass dependence on the emitted cluster, we systematically study the half-lives of cluster radioactivity for 26 trans-lead nuclei using CPPM with 28 different versions of proximity potential formalisms, which are all known proximity potential formalisms at present. The calculated results indicate that Prox.77-12 and Prox.81 are the best two with the lowest root-mean-square deviation for the study of cluster radioactivity.

This article is organized as follows. The theoretical framework of CPPM with 28 different
proximity potential formalisms and the compared semi-empirical and/or empirical formulae are exhaustively introduced in Section \ref{sec:Theoretical framework}. The results and discussion are distinctly presented in Section \ref{sec:Results and discussion}. Finally, a brief summary is given in Section \ref{sec:Summary}.

\section{Theoretical framework}
\label{sec:Theoretical framework}
\subsection{The half-lives of the cluster radioactivity}
The cluster radioactivity half-life can be determined by \cite{2023CPC47014101}
\begin{eqnarray}\label{1}
{T}_{1/2} = \frac{\rm{ln2}}{\lambda}=\frac{\rm{ln2}}{\nu{S}_{c}{P}},
\end{eqnarray}
where $\lambda$ is the decay constant. $\nu$ is the assault frequency on the barrier per second, which is taken as $1.0\times10^{22} s^{-1}$ in this work \cite{1991PS44427, 2002PRC65054308, 2011PRC83014601}. ${S}_{c}$ is the penetrability of barrier internal part (equal to the preformation probability of the cluster at the nuclear surface in a $\alpha$-like theory). It has been suggested that in the case of heavy cluster radioactivity \cite{1988PRL611930}, ${S}_{c}$ can be expressed as
\begin{eqnarray}\label{2}
{S}_{c}=({S}_{\alpha})^{({A_{c}-1})/{3}},
\end{eqnarray}
where $A_{c}$ is the mass number of the cluster and ${S}_{\alpha}$ is the
preformation probability for the $\alpha$ decay. In different models, some investigators obtained similar ${S}_{\alpha}$ values by fitting the experimental data \cite{2009PRC80037307, 2009EPJA41197, 2008PRC77027603, 2021CPC45044111}. In this study, we choose  
${S}_{\alpha}=0.02897$ for even-even parent nuclei and ${S}_{\alpha}=0.0214$ for odd-A parent nuclei \cite{2009PRC80037307}. The cluster radioactivity penetration probability through the potential barrier $P$ can be calculated by using the semiclassical WKB approximation action integral and expressed as
\begin{eqnarray}\label{3}
P=\exp(-\frac{2}{\hbar}\int_{R_{in}}^{R_{out}}\sqrt{{2\mu}\left\lvert{V(r)-Q_{c}}\right\rvert}dr),
\end{eqnarray}
where $\hbar$ is the reduced Planck constant. $\mu=\frac{m_{d}m_{c}}{m_{d}+m_{c}}$ is the
reduced mass of emitted cluster-daughter nucleus system with $m_{d}$ and $m_{c}$ being the daughter nucleus and the emitted cluster mass, respectively \cite{2016NPA95186}. $Q_{c}$ represents the cluster radioactivity released energy. It can be obtained by using \cite{2015MPLA301550150}
\begin{eqnarray}\label{4}
Q_{c}=B(A_{c},Z_{c})+B(A_{d},Z_{d})-B(A_{p},Z_{p}),
\end{eqnarray} 
where $B(A_{c},Z_{c})$, $B(A_{d},Z_{d})$ and $B(A_{p},Z_{p})$ are the binding energy of the emitted cluster, daughter and parent nucleus, respectively. They are taken from AME2020 \cite{2021CPC45030003} and NUBASE2020 \cite{2021CPC45030001}. $A_{c}, Z_{c}$, $A_{d}, Z_{d}$ and $A_{p}, Z_{p}$ are the mass numbers and
proton numbers of the emitted cluster, daughter and parent nucleus, respectively.

The total interaction potential $V(r)$ between the emitted cluster and daughter nucleus is composed of the nuclear potential $V_{N}(r)$, Coulomb potential $V_{C}(r)$ and centrifugal potential $V_{\ell}(r)$. It can be expressed as
\begin{eqnarray}\label{5}
V(r)=V_{N}(r)+V_{C}(r)+V_{\ell}(r).
\end{eqnarray}
In this work, we adopt the proximity potential formalism to substitute nuclear potential $V_{N}(r)$, the detailed information is given in section \ref{The proximity potential formalism}.
The Coulomb potential $V_{C}(r)$ is taken as the potential of a uniformly charged
sphere with radius $R$, which can be expressed as
\begin{eqnarray}\label{6}
V_{C}(r)=\left\{\begin{array}{ll}
\frac{Z_{c}Z_{d}e^{2}}{2R}[3-(\frac{r}{R})^{2}],&r\leq R,\\
\frac{Z_{c}Z_{d}e^{2}}{r},&r>R,
\end{array}\right.
\end{eqnarray}
where $e^{2}=1.4399652$ MeV$\cdot$fm is the square of the electronic elementary charge. $R=R_{d}+R_{c}$ is the sharp radius with $R_{d}$ and $R_{c}$ being the radii of daughter nucleus and emitted cluster, respectively. Various expressions for $R_{i} (i=c,d)$ within different proximity potential formalisms are in the section \ref{The proximity potential formalism}.

For the centrifugal potential $V_{\ell}(r)$, we adopt the Langer modified form since ${\ell}({\ell+1})\to({\ell+\dfrac{1}{2}})^{2}$ is a necessary correction for one-dimensional problems \cite{1995JMP365431}. It can be written as
\begin{eqnarray}\label{7}
V_{\ell}(r)=\frac{\left(\ell+1/2\right)^{2}\hbar^{2}}{2\mu{r}^{2}},
\end{eqnarray}
where $\ell$ is the angular momentum carried by the emitted cluster, which can be obtained via \cite{2023CPC47094103}
\begin{eqnarray}\label{8}
\ell=\left\{\begin{array}{llll}
\Delta_{j}, &$for even $\Delta_{j}$ and $\pi_{p} = \pi_{d}$$,\\
\Delta_{j}+1, &$for even $\Delta_{j}$ and $\pi_{p} \neq \pi_{d}$$,\\
\Delta_{j}, &$for odd $\Delta_{j}$ and $\pi_{p} \neq \pi_{d}$$,\\
\Delta_{j}+1, &$for odd $\Delta_{j}$ and $\pi_{p} = \pi_{d}$$.
\end{array}\right.
\end{eqnarray}
Here $\Delta_{j}$ = $\lvert {j_{p}-j_{d}-j_{c}} \rvert $, $j_{c}$, $\pi_{c}$, $j_{d}$, $\pi_{d}$ and $j_{p}$, $\pi_{p}$ are the isospin and parity values of the emitted cluster, daughter and parent nuclei, respectively. They are taken from NUBASE2020 \cite{2021CPC45030001}. As for the bounds of integral in the equation (\ref{3}), $R_{in}=R_{d}+R_{c}$ and $R_{out}=\frac{Z_{c}Z_{d}e^{2}}{2Q_{c}}+\sqrt{(\frac{Z_{c}Z_{d}e^{2}}{2Q_{c}})^{2}+\frac{\hbar^{2}(l+1/2)^{2}}{2\mu Q_{c}}}$ are the radius for the separation configuration and the outer turning point \cite{2009EPJA41197}.
\subsection{The proximity potential formalism}
\label{The proximity potential formalism}
In the present work, we choose 28 versions of proximity
potential formalisms to calculate the emitted cluster-daughter nucleus nuclear potential $V_{N}(r)$, which are: (i) Prox.77 \cite{1977AP105427} and its 12 modified forms on the basis of adjusting the surface energy coefficient $\gamma_{0}$ and $k_{s}$ \cite{1967AF36343, 1966NP811, 1976NPA272502, 1979PRC20992, 1981NPA361117, 1984JPG101057, 1988ADNDT39213, 1995ADNDT59185, 2003PRC67044316}, (ii) Prox.81 \cite{1981AP13253}, (iii) Prox.00 \cite{2000PRC62044610} and its revised versions Prox.00DP \cite{2010PRC81044615}, Prox.2010 \cite{2010CPL27112402} and Dutt2011 \cite{2011P76921}, (iv) Bass73 \cite{1973PLB47139} and its revised versions Bass77 \cite{1977PRL39265} and Bass80 \cite{1994JPG201297}, (v) CW76 \cite{1976PLB6519} and its revised versions BW91 \cite{1994JPG201297} and AW95 \cite{1995NPA594203}, (vi) Ng$\hat{o}$80 \cite{1980NPA348140}, (vii) Denisov \cite{2002PLB526315} and its revised version Denisov DP \cite{2010PRC81044615}, (viii) Guo2013 \cite{2013NPA89754}. Their detailed expressions are listed in the following.
\subsubsection{ The proximity potential 77 family}
In 1970s, the original version of the proximity
potential formalism for two spherical interacting nuclei was proposed by Blocki \textit{et al.} \cite{1977AP105427}, which can be expressed as
\begin{eqnarray}\label{9}
V_{N}(r)=4\pi\gamma{b}\bar{R}\Phi(\xi).
\end{eqnarray}
Here $\gamma$ is the surface energy coefficient based on Myers and \'{S}wia\c{}tecki formula \cite{1967AF36343} and is written in the following forms
\begin{eqnarray}\label{10}
\gamma=\gamma_{0}(1-k_{s}I^{2}),
\end{eqnarray}
where $I=\frac{N_{p}-Z_{p}}{A_{p}}$ is the asymmetry parameter and refers
to neutron-proton excess of the parent nucleus with $N_{p}$, $Z_{p}$ and $A_{p}$ being the neutron, proton and mass numbers of the parent nucleus, respectively. $\gamma_{0}$ and $k_{s}$ are the surface energy constant and the surface asymmetry constant of Prox.77 and its modifications, respectively. Their details are listed in Table \ref{Tab1}. 

\begin{table}[!htb] 
	\renewcommand\arraystretch{1.4}
	\setlength{\tabcolsep}{1.8mm}
	\caption{The different sets of the surface energy coefficient. $\gamma_{0}$
		and $k_{s}$ are the surface energy constant and the surface asymmetry
		constant, respectively.}
	\label {Tab1}
	\footnotesize
	\begin{tabular*}{8cm} {@{\extracolsep{\fill}} lccc}
		\hline
		\hline
		$\gamma$ set             &$\gamma_{0}$(MeV/fm$^{2}$) &$k_{s}$ & References \\ 
		\hline
		Set1($\gamma$-MS 1967)   &0.9517                     &1.7826  & \cite{1967AF36343}\\
		Set2($\gamma$-MS 1966)   &1.01734                    &1.79    & \cite{1966NP811}\\
		Set3($\gamma$-MN 1976)   &1.460734                   &4.0     & \cite{1976NPA272502}\\
		Set4($\gamma$-KNS 1979)  &1.2402                     &3.0     & \cite{1979PRC20992}\\
		Set5($\gamma$-MN-I 1981) &1.1756                     &2.2     & \cite{1981NPA361117}\\
		Set6($\gamma$-MN-II 1981) &1.27326                     &2.5     & \cite{1981NPA361117}\\
		Set7($\gamma$-MN-III 1981) &1.2502                     &2.4     & \cite{1981NPA361117}\\
		Set8($\gamma$-RR 1984) &0.9517                     &2.6     & \cite{1984JPG101057}\\
		Set9($\gamma$-MN 1988) &1.2496                     &2.3     & \cite{1988ADNDT39213}\\
		Set10($\gamma$-MN 1995) &1.25284                     &2.345     & \cite{1995ADNDT59185}\\
		Set11($\gamma$-PD-LDM 2003) &1.08948                     &1.9830   & \cite{2003PRC67044316}\\
		Set12($\gamma$-PD-NLD 2003) &0.9180                     &0.7546    & \cite{2003PRC67044316}\\
		Set13($\gamma$-PD-LSD 2003) &0.911445                     &2.2938  & \cite{2003PRC67044316}\\
		\hline\hline
		\end{tabular*}  
\end{table} 

$\bar{R}$ is the mean curvature radius or reduced radius. It can be obtained by
\begin{eqnarray}\label{11}
\bar{R}=\frac{C_{c}C_{d}}{C_{c}+C_{d}},
\end{eqnarray}
where $C_{i}=R_{i}[1-(\frac{b}{R_{i}})^{2}]$ $(i=c,d)$ represents the matter radius of the emitted cluster $(i=c)$ and daughter nucleus $(i=d)$, respectively. $R_{i}=1.28A_{i}^{1/3}-0.76+0.8A_{i}^{-1/3}$ $(i=c,d)$ is the effective sharp radius with $A_{i}$ being the mass number of the emitted cluster $(i=c)$ and daughter nucleus $(i=d)$, respectively.
The diffuseness of nuclear surface \textit{b} is considered close to unity ($b\approx 1$ fm).
The universal function $\Phi(\xi)$ is expressed as
\begin{eqnarray}\label{12}
\begin{footnotesize}
\Phi(\xi)=\left\{\begin{array}{ll}
-\frac{1}{2}(\xi-2.54)^{2}-0.0852(\xi-2.54)^{3},&\xi<1.2511,\\
-3.437\exp(-\frac{\xi}{0.75}),&\xi\geq{1.2511},
\end{array}\right.
\end{footnotesize}
\end{eqnarray}
where $\xi=\frac{r-C_{c}-C_{d}}{b}$ is the distance between the near surface of the emitted cluster and daughter nucleus.
\subsubsection{ The proximity potential Prox.81}
In 1981, a new proximity potential formalism was proposed by Blocki and \'{S}wia\c{}tecki based on the proximity force theorem, which is labeled as Prox.81 \cite{1981AP13253}. It has the same form as Prox.77 \cite{1977AP105427} except for the surface energy coefficient $\gamma=0.9517[1-1.7826(\dfrac{N_{p}-Z_{p}}{A_{p}})^{2}]$ and the universal function $\Phi(\xi=\frac{r-C_{c}-C_{d}}{b})$ written as  
\begin{eqnarray}\label{13} 
\begin{footnotesize}
\Phi(\xi)=\left\{\begin{array}{lll}
-1.7817+0.9270\xi+0.143\xi^{2}-0.09\xi^{3},\quad\xi<0,\\
-1.7817+0.9270\xi+0.01696\xi^{2}\\
-0.05148\xi^{3},\qquad\qquad\qquad\qquad 0\leq\xi\leq{1.9475},\\
-4.41\exp(-\frac{\xi}{0.7176}),\qquad\qquad\qquad\ \xi>{1.9475}.
\end{array}\right.
\end{footnotesize}
\end{eqnarray}
\subsubsection{ The proximity potential Prox.00}
In 2000, Myers \textit{et al.} proposed a fresh
proximity potential formalism to study the cross sections of synthesis new superheavy nuclei \cite{2000PRC62044610}. It is labeled as Prox.00 and expressed as
\begin{eqnarray}\label{14}
V_{N}(r)=4\pi\gamma{b}\bar{R}\Phi(\xi),
\end{eqnarray}
where $b$ is the width parameter taken as unity. $\gamma$ is the surface energy coefficient given by
\begin{eqnarray}\label{15}
\gamma=\frac{1}{4\pi{r_{0}^{2}}}{[18.63-Q\frac{(t_{c}^{2}+t_{d}^{2})}{2r_{0}^{2}}]},
\end{eqnarray}
with neutron skin of nucleus
\begin{eqnarray}\label{16}
t_{i}=\frac{3}{2}r_{0}[\frac{J(\frac{N_{i}-Z_{i}}{A_{i}})-\frac{1}{12}gZ_{i}A_{i}^{-1/3}}{Q+\frac{9}{4}A_{i}^{-1/3}}],&(i=c,d).
\end{eqnarray}
Here $r_{0}=1.14$ fm, the nuclear symmetric energy coefficient $J=32.65$ MeV and
$g=0.757895$ MeV, the neutron skin stiffness coefficient $Q=35.4$ MeV \cite{2000PRC62044610}. $N_{i}$, $Z_{i}$ and $A_{i}$ $(i=c, d)$ refer to the neutron, proton and mass numbers of the emitted cluster and
daughter nuclei, respectively. $\bar{R}=\frac{C_{c}C_{d}}{C_{c}+C_{d}}$ is the mean curvature radius. $C_{c}$ and $C_{d}$ are the matter radius of the emitted cluster and daughter nucleus, which can expressed as
\begin{eqnarray}\label{17}
C_{i}=c_{i}+\frac{N_{i}}{A_{i}}t_{i},&(i=c,d),
\end{eqnarray}
with the half-density radius of the charge distribution
\begin{eqnarray}\label{18}
c_{i}=R_{i}(1-\frac{7b^{2}}{R_{i}^{2}}-\frac{49b^{4}}{8R_{i}^{4}}),&(i=c,d).
\end{eqnarray}
Here $R_{i}$ is the nuclear charge radius which can be expressed as
\begin{eqnarray}\label{19}
R_{i}=1.256A_{i}^{1/3}[1-0.202(\frac{N_{i}-Z_{i}}{A_{i}})],&(i=c,d).
\end{eqnarray}
The universal function $\Phi(\xi=\frac{r-C_{c}-C_{d}}{b})$ is expressed as
\begin{eqnarray}\label{20}
\begin{footnotesize}
\Phi(\xi)=\left\{\begin{array}{ll}
-0.1353+\sum_{n=0}^{5}\frac{c_{n}}{n+1}(2.5-\xi)^{n+1},&0<\xi<2.5,\\
-0.9551\exp(\frac{2.75-\xi}{0.7176}),&\xi\geq{2.5},
\end{array}\right.
\end{footnotesize}
\end{eqnarray}
where the values of different constants $c_{n}$ are: $c_{0}=-0.1886$, $c_{1}=-0.2628$, $c_{2}= -0.15216$, $c_{3}=-0.04562$, $c_{4}=-0.069136$ and $c_{5}=-0.011454$ \cite{2000PRC62044610}.
\subsubsection{ The proximity potential Prox.00 DP}
A modified version of Prox.00 was proposed by Dutt \textit{et al.} using a more precise radius formula given by Royer and Roisseau \cite{2009EPJA42541}, which is labeled as Prox.00 DP \cite{2010PRC81044615}. It is same as Prox.00 except for the neutron skin of the emitted cluster $t_{c}=0$ and the nuclear charge radius $R_{i}$ written as
\begin{eqnarray}\label{21}
R_{i}=1.2332A_{i}^{1/3}+\frac{2.8961}{A_{i}^{2/3}}\nonumber\\
-0.18688A_{i}^{1/3}\frac{N_{i}-Z_{i}}{A_{i}},&(i=c,d).
\end{eqnarray}
\subsubsection{ The proximity potential Prox.2010}
In 2010, Dutt and Bensal presented another modified version of Prox.00 denoted as Prox.2010 \cite{2010CPL27112402}. It has the same form as Prox.00 DP except for the surface energy coefficient $\gamma=1.25284[1-2.345(\dfrac{N_{p}-Z_{p}}{A_{p}})^{2}]$ and the universal function $\Phi(\xi=\frac{r-C_{c}-C_{d}}{b})$ written as 
\begin{eqnarray}\label{22} 
\begin{footnotesize}
\Phi(\xi)=\left\{\begin{array}{lll}
-1.7817+0.9270\xi+0.143\xi^{2}-0.09\xi^{3},\quad\xi<0,\\
-1.7817+0.9270\xi+0.01696\xi^{2}\\
-0.05148\xi^{3},\qquad\qquad\qquad\qquad 0\leq\xi\leq{1.9475},\\
-4.41\exp(-\frac{\xi}{0.7176}),\qquad\qquad\qquad\ \xi>{1.9475}.
\end{array}\right.
\end{footnotesize}
\end{eqnarray}
\subsubsection{ The proximity potential Dutt2011}
Based on the Prox.77, Dutt presented a new version of potential, which is labeled as Dutt2011  \cite{2011P76921}.
It has the same form as Prox.2010 except for the nuclear charge radius $R_{i}$ written as 
\begin{eqnarray}\label{23}
R_{i}=1.171A_{i}^{1/3}+1.427A_{i}^{-1/3},&(i=c,d).
\end{eqnarray}
\subsubsection{ The proximity potential Bass73}
In 1973, Bass obtained the nuclear potential with the difference in surface energies
between finite and infinite separation based on the liquid drop model \cite{1973PLB47139}. It is labeled as Bass73 and expressed as
\begin{eqnarray}\label{24}
V_{N}(r)=\frac{-da_{s}A_{c}^{1/3}A_{d}^{1/3}}{R}\exp(-\frac{r-R}{d}),
\end{eqnarray}
where $R=R_{c}+R_{d}=r_{0}(A_{c}^{1/3}+A_{d}^{1/3})$ is the sum of the half-maximum density radii, where $r_{0}=1.07$ fm, $R_{c}$, $A_{c}$, and $R_{d}$, $A_{d}$ are the radii
and mass numbers of daughter nucleus and emitted cluster, respectively. $a_{s}=17.0$ MeV and $d=1.35$ fm are the surface term in the liquid drop model mass formula and the range parameter, respectively.
\subsubsection{ The proximity potential Bass77}
For the Bass77, the nuclear potential $V_{N}$ is given by \cite{1977PRL39265}
\begin{eqnarray}\label{25}
V_{N}(r)=-\frac{R_{c}R_{d}}{R_{c}+R_{d}}\Phi(s),
\end{eqnarray}
where $R_{i}=1.16A_{i}^{1/3}-1.39A_{i}^{-1/3}$ $(i=c, d)$ is the half-density radius with $A_{i}$ being the mass number of the emitted cluster $(i=c)$ and daughter nucleus $(i=d)$, respectively. The universal function $\Phi(s=r-R_{c}-R_{d})$ can be given by
\begin{eqnarray}\label{26}
\Phi(s)=[0.03\exp(\frac{s}{3.3})+0.0061\exp(\frac{s}{0.65})]^{-1}.
\end{eqnarray}
\subsubsection{ The proximity potential Bass80}
Based on the proximity potential Bass77, Bass proposed an improved proximity potential formalism, which is labeled as Bass80 \cite{1994JPG201297}. It is expressed as
\begin{eqnarray}\label{27}
V_{N}(r)=-\frac{R_{c}R_{d}}{R_{c}+R_{d}}\Phi(s).
\end{eqnarray}
Here $R_{i}=R_{si}(1-\frac{0.98}{R_{si}^{2}})$ with $R_{si}=1.28A_{i}^{1/3}-0.76+0.8A_{i}^{-1/3}$ $(i=c, d)$. The universal function $\Phi(s=r-R_{c}-R_{d})$ is written by
\begin{eqnarray}\label{28}
\Phi(s)=[0.033\exp(\frac{s}{3.5})+0.007\exp(\frac{s}{0.65})]^{-1}.
\end{eqnarray}
\subsubsection{ The proximity potential CW76}
In 1976, an empirical nuclear potential was proposed by Christensen and Winter based on the analysis of heavy-ion elastic scattering data \cite{1976PLB6519}. It is labeled as CW76 and expressed as
\begin{eqnarray}\label{29}
V_{N}(r)=-50\frac{R_{c}R_{d}}{R_{c}+R_{d}}\Phi(s),
\end{eqnarray}
where $R_{c}$ and $R_{d}$ are given by
\begin{eqnarray}\label{30}
R_{i}=1.233A_{i}^{1/3}-0.978A_{i}^{-1/3}, &(i=c,d).
\end{eqnarray}
The universal function $\Phi(s=r-R_{c}-R_{d})$ has the following form
\begin{eqnarray}\label{31}
\Phi(s)=\exp(-\frac{s}{0.63}).
\end{eqnarray}
\subsubsection{ The proximity potential BW91}
In 1991, Broglia and Winther presented a more refined nuclear potential by taking the Woods-Saxon parameterization of the proximity potential CW76. It is labeled as BW91 and expressed as \cite{1994JPG201297}
\begin{eqnarray}\label{32}
V_{N}(r)=-\frac{V_{0}}{1+\exp(\frac{r-R}{0.63})}
=-\frac{16\pi\gamma{a}\frac{R_{c}R_{d}}{R_{c}+R_{d}}}{1+\exp(\frac{r-R}{0.63})}.
\end{eqnarray}
Here $\gamma=0.95[1-1.8(\frac{N_{d}-Z_{d}}{A_{d}})(\frac{N_{c}-Z_{c}}{A_{c}})]$ is the surface energy coefficient with $N_{d}$, $Z_{d}$, $A_{d}$ and $N_{c}$, $Z_{c}$, $A_{c}$ being the neutron, proton and mass numbers of the daughter nucleus and emitted cluster, respectively. $a=0.63$ fm and $R=R_{c}+R_{d}+0.29$ with $R_{i}=1.233A_{i}^{1/3}-0.98A_{i}^{-1/3} (i=c,d)$. 
\subsubsection{ The proximity potential AW95}
For the Aage Withner(AW95) potential \cite{1995NPA594203}, the proximity potential expression and other parameters are the same as BW91, except for
\begin{eqnarray}\label{33}
a=\frac{1}{1.17(1+0.53(A_{c}^{-1/3}+A_{d}^{-1/3}))}
\end{eqnarray}
and $R=R_{c}+R_{d}$ with $R_{i}=1.2A_{i}^{1/3}-0.09$ $(i=c,d)$.
\subsubsection{ The proximity potential Ng$\hat{o}$80}
In 1980, H. Ng$\hat{o}$ and Ch. Ng$\hat{o}$ obtained the nuclear potential part of the interaction potential between two heavy ions using the energy density formalism and Fermi distributions for the nuclear densities \cite{1980NPA348140}. It is labeled as Ng$\hat{o}$ and expressed as
\begin{eqnarray}\label{34}
V_{N}(r)=\frac{C_{c}C_{d}}{C_{c}+C_{d}}\phi(\xi),
\end{eqnarray}
where $C_{i}=R_{i}[1-(\frac{b}{R_{i}})^{2}]$ $(i=c,d)$ is the S$\ddot{u}$smann central radii of the emitted cluster and daughter nucleus. $b$ is the diffuseness of nuclear surface taken as unity.
$R_{i}$ is the sharp radii and expressed as
\begin{eqnarray}\label{35}
R_{i}=\frac{N_{i}R_{ni}+Z_{i}R_{pi}}{A_{i}},&(i=c,d),
\end{eqnarray}
where $R_{ji}=r_{0ji}A_{i}^{1/3}$ $(j=p,n$, $i=c,d)$ with $r_{0pi}=1.128$ fm and $r_{0ni}=1.1375+1.875\times10^{-4}A_{i}$ fm. The universal function $\phi(\xi=r-C_{c}-C_{d})$ is given by
\begin{eqnarray}\label{36}
\Phi(\xi)=\left\{\begin{array}{ll}
-33+5.4(\xi+1.6)^{2},&\xi<-1.6,\\
-33\exp(-\frac{1}{5}(s+1.6)^{2}),&\xi\geq{-1.6}.
\end{array}\right.
\end{eqnarray}
\subsubsection{ The proximity potential Denisov}
By choosing 119 spherical or near spherical even-even nuclei around the $\beta$-stability line, Denisov presented a simple analytical expression for the nuclear potential of ion-ion interaction potential using the semi-microscopic approximation between all possible nucleus-nucleus combinations \cite{2002PLB526315}. It is labeled as Denisov and expressed as
\begin{eqnarray}\label{37}
V_{N}(r)=-1.989843\frac{R_{c}R_{d}}{R_{c}+R_{d}}\Phi(s)\times[1+0.003525139\nonumber\\
\times(\frac{A_{c}}{A_{d}}+\frac{A_{d}}{A_{c}})^{3/2}-0.4113263(I_{c}+I_{d})],
\end{eqnarray}
where $I_{i}=\frac{N_{i}-Z_{i}}{A_{i}}$ $(i=c,d)$ is the isospin asymmetry. $R_{i}$ is the effective nuclear radius and can be obtained by
\begin{eqnarray}\label{38}
R_{i}=R_{0i}(1-\frac{3.413817}{R_{0i}}^{2})+\nonumber\\
\quad\quad{1.284589(I_{i}-\frac{0.4A_{i}}{A_{i}+200})},&(i=c,d),
\end{eqnarray}
with
\begin{eqnarray}\label{39}
R_{0i}=1.240A_{i}^{1/3}(1+\frac{1.646}{A_{i}}-0.191I_{i}),&(i=c,d).
\end{eqnarray}
The universal function $\Phi(s=r-R_{c}-R_{d}-2.65)$ is expressed as
\begin{eqnarray}\label{40}
\begin{footnotesize}
\Phi(s)=\left\{\begin{array}{lr}
\{1-s^{2}[0.05410106\frac{R_{c}R_{d}}{R_{c}+R_{d}}\exp(-\frac{s}{1.760580})\\
-0.5395420(I_{c}+I_{d})\exp(-\frac{s}{2.424408})]\}\\
\times\exp(-\frac{s}{0.7881663}),\qquad\qquad\quad\qquad\qquad{s\geq{0}},\\
1-\frac{s}{0.7881663}+1.229218s^{2}-0.2234277s^{3}\\
-0.1038769s^{4}-\frac{R_{c}R_{d}}{R_{c}+R_{d}}(0.1844935s^{2}\\
+0.07570101s^{3})+(I_{c}+I_{d})\\
(0.04470645s^{2}+0.03346870s^{3}),-5.65\leq{s}\leq{0}.
\end{array}\right.
\end{footnotesize}
\end{eqnarray}
\subsubsection{ The proximity potential Denisov DP}
The proximity potential Denisov DP \cite{2010PRC81044615} is the modified version of Denisov using a more precise radius formula proposed by Royer \textit{et al.} \cite{2009EPJA42541} and expressed as
\begin{eqnarray}\label{41}
R_{i}=1.2332A_{i}^{1/3}+\frac{2.8961}{A_{i}^{2/3}}\nonumber\\
\qquad\qquad-0.18688A_{i}^{1/3}\frac{N_{i}-Z_{i}}{A_{i}},&(i=c,d).
\end{eqnarray}
\subsubsection{ The proximity potential Guo2013}
In 2013, Guo \textit{et al.} presented a universal function of nuclear proximity
potential from density-dependent nucleon-nucleon interaction using the double folding model \cite{2013NPA89754}. It is labeled as Guo2013 and expressed as
\begin{eqnarray}\label{42}
V_{N}(r)=4\pi\gamma{b}\frac{R_{c}R_{d}}{R_{c}+R_{d}}\Phi(s),
\end{eqnarray}
where $\gamma=0.9517[1-1.7826(\dfrac{N_{p}-Z_{p}}{A_{p}})^{2}]$ is the surface cofficient and $R_{i}=1.28A_{i}^{1/3}-0.76+0.8A_{i}^{-1/3}$ $(i=c,d)$ is the effective sharp radius.
The universal function $\Phi(s=\frac{r-R_{c}-R_{d}}{b})$ is expressed as
\begin{eqnarray}\label{43}
\Phi(s)=\frac{p_{1}}{1+\exp(\frac{s+p_{2}}{p_{3}})},
\end{eqnarray}
where $p_{1}=-17.72$, $p_{2}=1.30$ and $p_{3}=0.854$ are the adjustable parameters.

\subsection{Empirical and semi-empirical formulas}
\subsubsection{Universal decay law}
In 2009, a linear expression for charged-particle emissions was proposed by Qi \textit{et al.} based on $\alpha$-like $R$-matrix theory and named as UDL \cite{2009PRC80044326, 2009PRL103072501}. It can be expressed as
\begin{footnotesize}
	\begin{eqnarray}\label{44}
	\log_{10}T_{1/2}=aZ_{c}Z_{d}\sqrt{\frac{\mathcal{U}}{Q_{c}}}+b\sqrt{\mathcal{U}Z_{c}Z_{d}(A_{c}^{1/3}+A_{d}^{1/3})}+c,
	\end{eqnarray}
\end{footnotesize}where $Q_{c}$ represents the cluster radioactivity released energy. $\mathcal{U}=A_{c}A_{d}/(A_{c}+A_{d})$ is the reduced mass of the emitted cluster-daughter nucleus system measured in unit of the nucleon mass with $Z_{d}$, $A_{d}$ and $Z_{c}$,  $A_{c}$ being the proton and mass number of daughter nucleus and emitted cluster, respectively. $a=0.4314$, $b=-0.3921$ and $c=-32.7044$ are the adjustable parameters \cite{2009PRC80044326, 2009PRL103072501}.
\subsubsection{Ni's empirical formula}
In 2008, Ni \textit{et al.} proposed a unified formula of half-lives for $\alpha$ decay and cluster radioactivity by deducing from the WKB barrier penetration probability with some approximations \cite{2008PRC78044310}. It can be expressed as
\begin{eqnarray}\label{45}
\log_{10}T_{1/2}=\textit{a}\sqrt{\mathcal{U}}Z_{c}Z_{d}Q_{c}^{-1/2}+b\sqrt{\mathcal{U}}(Z_{c}Z_{d})^{1/2}+c,
\end{eqnarray}
where $Q_{c}$ represents $\alpha$ decay energy and cluster radioactivity released energy. $\mathcal{U}=A_{c}A_{d}/(A_{c}+A_{d})$ is the reduced mass the same as UDL. For cluster radioactivity, the adjustable parameters $a=0.38617$, $b=-1.08676$, $c=-21.37195$ for even-even nuclei and $c=-20.11223$ for odd-A nuclei \cite{2008PRC78044310}. For $\alpha$ decay, the adjustable parameters $a=0.39961$, $b=-1.31008$, $c=-17.00698$ for even-even nuclei, $c=-16.26029$ for even-odd nuclei and $c=-16.40484$ for odd-even nuclei \cite{2008PRC78044310}. 

\subsubsection{Scaling law}
In 2004, Horoi \textit{et al.} proposed the first model-independent scaling law to describe the regularities of the experimental data for cluster radioactivity \cite{2004JPG30945}, which can be expressed as
\begin{eqnarray}\label{46}
\log_{10}T_{1/2}=(a\mathcal{U}^{x}+b)[\frac{({Z_{c}Z_{d}})^{y}}{Q_{c}^{1/2}}-7]+(c\mathcal{U}^{x}+d),    
\end{eqnarray} 
where $Q_{c}$ and $\mathcal{U}=\frac{A_{d}A_{c}}{A_{d}+A_{c}}$ are the same as UDL. $a=9.1$, $b=-10.2,$, $c=7.39$, $d=-23.2$, $x=0.416$ and $y=0.613$ are the adjustable parameters \cite{2004JPG30945}.

\section{Results and discussion}
\label{sec:Results and discussion}
The main purpose of this work is to perform a comparative study of various proximity potential formalisms when they are applied to cluster radioactivity. In order to explore the most suitable proximity potential formalism for the cluster radioactivity, we systematically calculate the cluster radioactivity half-lives of 26 nuclei in the emission of clusters $^{14}$$\rm{C}$, $^{20}$$\rm{O}$, $^{23}$$\rm{F}$, $^{24,25,26}$$\rm{Ne}$, $^{28,30}$$\rm{Mg}$ and $^{32,34}$$\rm{Si}$ from various parent nuclei $^{221}$$\rm{Fr}$ to $^{242}$$\rm{Cm}$ by using CPPM with 28 different versions of proximity potential formalisms. In addition, a universal decay law (UDL), Ni's empirical formula, a scaling law (SL) are also used. The calculated results and experimental data are detailedly listed in Table \ref{Tab2}. 
In this table, the first to third columns are the cluster decay process, the cluster radioactivity released energy $Q_{c}$ and the angular momentum $\ell$ taken away by the emitted cluster, respectively. As for the last nine columns, they represent the experimental data of the cluster radioactivity half-lifes and the calculated ones obtained by using CPPM with 28 different versions of proximity potential formalisms, UDL, Ni's empirical formula and SL in logarithmic form, respectively. 
From this table, we can find that the results calculated by using Prox.77 and its 12 modified forms as well as Prox.81 and Ng$\hat{o}$80 are within an order of magnitude as the experimental data on the whole, which indicates the experimental data can faultlessly be reproduced. For Prox.00 and its revised versions Prox.00 DP, Prox.2010 and Dutt2011 as well as Guo2013, their calculations differ overall from the experimental data by two to four orders of magnitude. While as for Bass73 and its revised versions Bass77 and Bass80 as well as CW76 and its revised versions BW91 and AW95, the calculated values of the improved version are closer to the experimental data than those of the previous version. For Denisov, its calculated results are more about one to three order magnitude than the experimental data. However, the calculated results of its revised version Denisov DP are less about six to eleven order magnitude than the experimental data and less about seven to thirteen order magnitude than the Denisov. Its reduced order magnitude increases with the size of the cluster particle.

\begin{figure}[h]\centering
	\includegraphics[width=9cm]{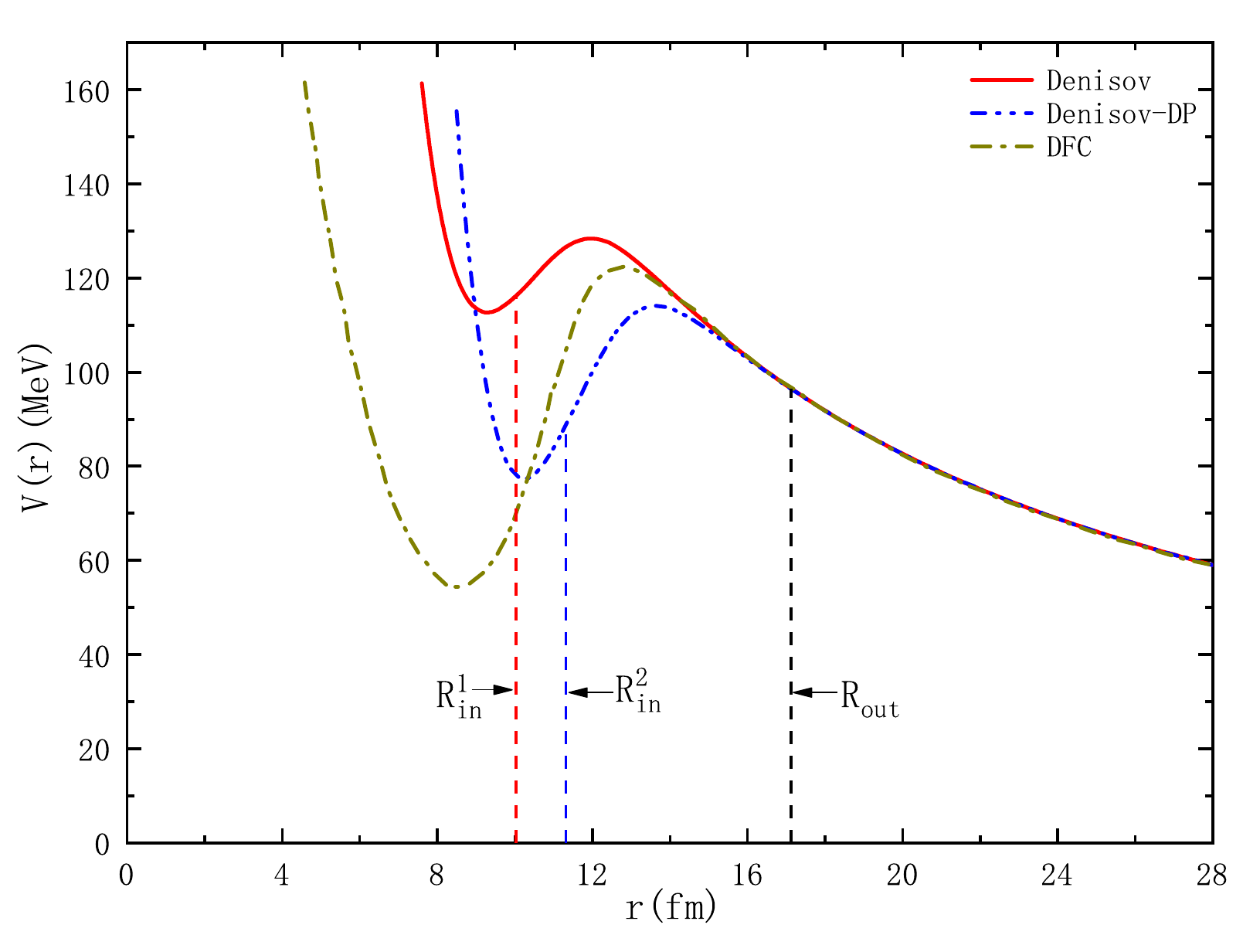}
	\caption{(color online) Schematic diagram of the total interaction potential $V(r)$ between the cluster and daughter nucleus using CPPM with Denisov and Denisov DP as well as DFC.
		The relevant data on DFC are obtained from Ref. \cite{2022ADNDT145101501}.	
		$R_{in}^{1}$ and $R_{in}^{2}$ are the radii for the separation configuration using Denisov and Denisov DP, respectively. $R_{out}$ is the outer turning point.}
	\label {fig 1}
\end{figure}

In order to explore the specific reasons for this situation, taking the example of $^{242}\rm{Cm}\to^{208}\rm{Pb}+^{34}\rm{Si}$, we plot the total interaction potential $V(r)$ between the emitted cluster and daughter nucleus using CPPM with Denisov and Denisov DP in Fig. \ref{fig 1}. In addition, the nuclear potential calculated by the double folding approach is also used in order to deeperly comprehend the difference between the proximity potential and other nuclear potentials, which takes into account the nuclear density distributions and the effective nucleon-nucleon interactions. It also has been shown to be successfully applied to $\alpha$ decay \cite{2020PRC102014327}, proton emission \cite{2016EPJA5268}, two-proton radioactivity \cite{2023SCICPMA66222012} and cluster radioactivity \cite{2016PRC94024315} in the density-dependent cluster model. In this work, we choose the monopole component of a realistic double folding potential plus Coulomb core-cluster potential (DFC) as the interaction potential to describe the cluster radioactivity. Its related data are taken from Ref. \cite{2022ADNDT145101501} and are also plotted in Fig. \ref{fig 1}. From this figure, we can find that the total interaction potential $V(r)$ at short distances changes dramatically, which shows the nuclear potential plays a dominating role whereas the choice of different nuclear potentials may lead to enormous differences. At long distances, $V(r)$ remains essentially constant, which should be mainly the effect of the Coulomb potential. In general, the trend of the potential energy curve remains consistent.
At the same time, the separation configuration radius $R_{in}$ of the improved Denisov DP is slightly increased compared with the original Denisov, but the outer turning point $R_{out}$ remains unchanged. Therefore, the integral result becomes smaller and the penetration probability $P$ is greater in Eq. (\ref{3}), which will result in a smaller cluster radioactivity half-life. This is consistent with the conclusions in the Ref. \cite{2016NPA95186}. For the larger cluster particles, the more cluster radioactivity released energy $Q_{c}$ is required that will give rise to the smaller outer turning point $R_{out}$, which will further reduce the cluster radioactivity half-life. Consequently, there is such a huge deviation between the calculated results of Denisov and Denisov DP, which is further expanded in the larger cluster particles.

\begingroup
\renewcommand*{\arraystretch}{1.5}
\setlength{\tabcolsep}{0.5mm}
\setlength\LTleft{0pt}
\setlength\LTright{0pt}
\setlength{\LTcapwidth}{7in}
\begin{longtable*}
	{@{\extracolsep{\fill}} cccccccccccc}
			\caption{Comparison of the discrepancy between the experimental cluster radioactivity half-lives (in seconds) and the calculated ones by using CPPM with 28 different versions of the proximity potential formalisms, UDL, Ni's empirical formula and SL in logarithmic form. The experimental cluster radioactivity half-lives are taken from Ref. \cite{2023CPC47064107}.}
			\label{Tab2}\\
			\hline 
			\hline			                       
			\endfirsthead
			\multicolumn{12}{c}
			{{\tablename\ \thetable{} -- continued from previous page}} \\
			\hline
			\endhead
			\hline \multicolumn{12}{r}{{Continued on next page}}\\
			\endfoot
			\endlastfoot
			{Cluster}&$Q_{c}$&\ $\ell$ \ &&&&&&$\rm{log}_{10}{\emph{T}}_{1/2}(s)$&&&\\
			\cline{4-12}
			decay&(MeV)&&\ EXP& Prox.77-1&Prox.77-2&Prox.77-3&Prox.77-4&Prox.77-5&Prox.77-6&Prox.77-7&Prox.77-8\\
			\hline          
			$^{221}$Fr$\to^{207}$Tl$+^{14}$C$$&$	31.29 	$&$	3	$&$	14.56 	$&$	14.82 	$&$	14.71 	$&$	14.17 	$&$	14.43 	$&$	14.46 	$&$	14.31 	$&$	14.35 	$&$	14.89 	$\\
			$^{221}$Ra$\to^{207}$Pb$+^{14}$C$$&$	32.40 	$&$	3	$&$	13.39 	$&$	13.69 	$&$	13.58 	$&$	12.99 	$&$	13.28 	$&$	13.32 	$&$	13.17 	$&$	13.20 	$&$	13.76 	$\\
			$^{222}$Ra$\to^{208}$Pb$+^{14}$C$$&$	33.05 	$&$	0	$&$	11.22 	$&$	11.80 	$&$	11.68 	$&$	11.09 	$&$	11.39 	$&$	11.42 	$&$	11.26 	$&$	11.30 	$&$	11.86 	$\\
			$^{223}$Ra$\to^{209}$Pb$+^{14}$C$$&$	31.83 	$&$	4	$&$	15.05 	$&$	14.72 	$&$	14.61 	$&$	14.06 	$&$	14.33 	$&$	14.36 	$&$	14.21 	$&$	14.24 	$&$	14.79 	$\\
			$^{224}$Ra$\to^{210}$Pb$+^{14}$C$$&$	30.53 	$&$	0	$&$	15.87 	$&$	16.39 	$&$	16.27 	$&$	15.75 	$&$	16.01 	$&$	16.03 	$&$	15.89 	$&$	15.92 	$&$	16.45 	$\\
			$^{226}$Ra$\to^{212}$Pb$+^{14}$C$$&$	28.20 	$&$	0	$&$	21.20 	$&$	21.28 	$&$	21.17 	$&$	20.70 	$&$	20.93 	$&$	20.95 	$&$	20.81 	$&$	20.84 	$&$	21.34 	$\\
			$^{223}$Ac$\to^{209}$Bi$+^{14}$C$$&$	33.06 	$&$	2	$&$	12.60 	$&$	13.23 	$&$	13.12 	$&$	12.52 	$&$	12.82 	$&$	12.86 	$&$	12.70 	$&$	12.73 	$&$	13.29 	$\\
			$^{225}$Ac$\to^{211}$Bi$+^{14}$C$$&$	30.48 	$&$	4	$&$	17.16 	$&$	18.18 	$&$	18.07 	$&$	17.54 	$&$	17.80 	$&$	17.83 	$&$	17.69 	$&$	17.72 	$&$	18.24 	$\\
			$^{228}$Th$\to^{208}$Pb$+^{20}$O$$&$	44.72 	$&$	0	$&$	20.73 	$&$	21.73 	$&$	21.59 	$&$	20.89 	$&$	21.23 	$&$	21.27 	$&$	21.08 	$&$	21.12 	$&$	21.82 	$\\
			$^{231}$Pa$\to^{208}$Pb$+^{23}$F$$&$	51.88 	$&$	1	$&$	26.02 	$&$	25.20 	$&$	25.04 	$&$	24.28 	$&$	24.66 	$&$	24.69 	$&$	24.49 	$&$	24.53 	$&$	25.29 	$\\
			$^{230}$Th$\to^{206}$Hg$+^{24}$Ne$$&$	57.76 	$&$	0	$&$	24.63 	$&$	24.52 	$&$	24.36 	$&$	23.63 	$&$	23.99 	$&$	24.02 	$&$	23.82 	$&$	23.86 	$&$	24.61 	$\\
			$^{231}$Pa$\to^{207}$Tl$+^{24}$Ne$$&$	60.41 	$&$	1	$&$	22.89 	$&$	22.91 	$&$	22.74 	$&$	21.96 	$&$	22.35 	$&$	22.39 	$&$	22.18 	$&$	22.22 	$&$	23.00 	$\\
			$^{232}$U$\to^{208}$Pb$+^{24}$Ne$$&$	62.31 	$&$	0	$&$	20.39 	$&$	20.45 	$&$	20.28 	$&$	19.45 	$&$	19.86 	$&$	19.91 	$&$	19.69 	$&$	19.74 	$&$	20.54 	$\\
			$^{233}$U$\to^{209}$Pb$+^{24}$Ne$$&$	60.49 	$&$	2	$&$	24.84 	$&$	23.95 	$&$	23.79 	$&$	23.01 	$&$	23.40 	$&$	23.44 	$&$	23.23 	$&$	23.27 	$&$	24.04 	$\\
			$^{234}$U$\to^{210}$Pb$+^{24}$Ne$$&$	58.83 	$&$	0	$&$	25.93 	$&$	25.26 	$&$	25.11 	$&$	24.37 	$&$	24.73 	$&$	24.77 	$&$	24.57 	$&$	24.61 	$&$	25.36 	$\\
			$^{235}$U$\to^{211}$Pb$+^{24}$Ne$$&$	57.36 	$&$	1	$&$	27.42 	$&$	28.46 	$&$	28.31 	$&$	27.61 	$&$	27.95 	$&$	27.98 	$&$	27.78 	$&$	27.83 	$&$	28.55 	$\\
			$^{233}$U$\to^{208}$Pb$+^{25}$Ne$$&$	60.70 	$&$	2	$&$	24.84 	$&$	24.51 	$&$	24.34 	$&$	23.52 	$&$	23.93 	$&$	23.97 	$&$	23.75 	$&$	23.80 	$&$	24.61 	$\\
			$^{234}$U$\to^{208}$Pb$+^{26}$Ne$$&$	59.41 	$&$	0	$&$	25.93 	$&$	26.10 	$&$	25.93 	$&$	25.11 	$&$	25.51 	$&$	25.55 	$&$	25.33 	$&$	25.38 	$&$	26.20 	$\\
			$^{234}$U$\to^{206}$Hg$+^{28}$Mg$$&$	74.11 	$&$	0	$&$	25.53 	$&$	24.90 	$&$	24.72 	$&$	23.90 	$&$	24.30 	$&$	24.34 	$&$	24.12 	$&$	24.16 	$&$	25.01 	$\\
			$^{236}$U$\to^{208}$Hg$+^{28}$Mg$$&$	70.73 	$&$	0	$&$	27.58 	$&$	29.28 	$&$	29.11 	$&$	28.36 	$&$	28.72 	$&$	28.75 	$&$	28.54 	$&$	28.59 	$&$	29.38 	$\\
			$^{236}$Pu$\to^{208}$Pb$+^{28}$Mg$$&$	79.67 	$&$	0	$&$	21.52 	$&$	20.70 	$&$	20.51 	$&$	19.57 	$&$	20.04 	$&$	20.10 	$&$	19.85 	$&$	19.90 	$&$	20.80 	$\\
			$^{238}$Pu$\to^{210}$Pb$+^{28}$Mg$$&$	75.91 	$&$	0	$&$	25.70 	$&$	25.10 	$&$	24.92 	$&$	24.08 	$&$	24.50 	$&$	24.54 	$&$	24.31 	$&$	24.36 	$&$	25.20 	$\\
			$^{236}$U$\to^{206}$Hg$+^{30}$Mg$$&$	72.27 	$&$	0	$&$	27.58 	$&$	28.72 	$&$	28.53 	$&$	27.70 	$&$	28.11 	$&$	28.14 	$&$	27.90 	$&$	27.95 	$&$	28.83 	$\\
			$^{238}$Pu$\to^{208}$Pb$+^{30}$Mg$$&$	76.79 	$&$	0	$&$	25.70 	$&$	25.45 	$&$	25.26 	$&$	24.33 	$&$	24.79 	$&$	24.84 	$&$	24.58 	$&$	24.64 	$&$	25.56 	$\\
			$^{238}$Pu$\to^{206}$Hg$+^{32}$Si$$&$	91.19 	$&$	0	$&$	25.28 	$&$	25.02 	$&$	24.82 	$&$	23.87 	$&$	24.34 	$&$	24.39 	$&$	24.13 	$&$	24.19 	$&$	25.13 	$\\
			$^{242}$Cm$\to^{208}$Pb$+^{34}$Si$$&$	96.54 	$&$	0	$&$	23.15 	$&$	23.01 	$&$	22.79 	$&$	21.69 	$&$	22.24 	$&$	22.30 	$&$	22.01 	$&$	22.07 	$&$	23.14 	$\\
			\hline
			{Cluster}&$Q_{c}$&\ $\ell$ \ &&&&&$\rm{log}_{10}{\emph{T}}_{1/2}(s)$&&&\\
			\cline{4-12}
			decay&(MeV)&&\ EXP& Prox.77-9&Prox.77-10&Prox.77-11&Prox.77-12&Prox.77-13&Prox.81&Prox.00&Prox.00 DP \\
			\hline
			$^{221}$Fr$\to^{207}$Tl$+^{14}$C$$&$	31.29 	$&$	3	$&$	14.56 	$&$	14.33 	$&$	14.33 	$&$	14.60 	$&$	14.80 	$&$	14.93 	$&$	14.75 	$&$	14.26 	$&$	11.87 	$\\
			$^{221}$Ra$\to^{207}$Pb$+^{14}$C$$&$	32.40 	$&$	3	$&$	13.39 	$&$	13.19 	$&$	13.19 	$&$	13.46 	$&$	13.68 	$&$	13.80 	$&$	13.62 	$&$	13.09 	$&$	10.72 	$\\
			$^{222}$Ra$\to^{208}$Pb$+^{14}$C$$&$	33.05 	$&$	0	$&$	11.22 	$&$	11.28 	$&$	11.28 	$&$	11.56 	$&$	11.78 	$&$	11.91 	$&$	11.73 	$&$	11.27 	$&$	8.91 	$\\
			$^{223}$Ra$\to^{209}$Pb$+^{14}$C$$&$	31.83 	$&$	4	$&$	15.05 	$&$	14.23 	$&$	14.23 	$&$	14.50 	$&$	14.70 	$&$	14.83 	$&$	14.65 	$&$	14.14 	$&$	11.74 	$\\
			$^{224}$Ra$\to^{210}$Pb$+^{14}$C$$&$	30.53 	$&$	0	$&$	15.87 	$&$	15.91 	$&$	15.91 	$&$	16.17 	$&$	16.36 	$&$	16.49 	$&$	16.32 	$&$	15.78 	$&$	13.36 	$\\
			$^{226}$Ra$\to^{212}$Pb$+^{14}$C$$&$	28.20 	$&$	0	$&$	21.20 	$&$	20.83 	$&$	20.83 	$&$	21.07 	$&$	21.25 	$&$	21.38 	$&$	21.21 	$&$	20.62 	$&$	18.14 	$\\
			$^{223}$Ac$\to^{209}$Bi$+^{14}$C$$&$	33.06 	$&$	2	$&$	12.60 	$&$	12.72 	$&$	12.72 	$&$	13.00 	$&$	13.22 	$&$	13.34 	$&$	13.16 	$&$	12.63 	$&$	10.27 	$\\
			$^{225}$Ac$\to^{211}$Bi$+^{14}$C$$&$	30.48 	$&$	4	$&$	17.16 	$&$	17.71 	$&$	17.71 	$&$	17.96 	$&$	18.16 	$&$	18.28 	$&$	18.11 	$&$	17.50 	$&$	15.06 	$\\
			$^{228}$Th$\to^{208}$Pb$+^{20}$O$$&$	44.72 	$&$	0	$&$	20.73 	$&$	21.11 	$&$	21.11 	$&$	21.44 	$&$	21.71 	$&$	21.87 	$&$	21.65 	$&$	20.78 	$&$	18.34 	$\\
			$^{231}$Pa$\to^{208}$Pb$+^{23}$F$$&$	51.88 	$&$	1	$&$	26.02 	$&$	24.52 	$&$	24.52 	$&$	24.88 	$&$	25.17 	$&$	25.35 	$&$	25.11 	$&$	24.05 	$&$	21.62 	$\\
			$^{230}$Th$\to^{206}$Hg$+^{24}$Ne$$&$	57.76 	$&$	0	$&$	24.63 	$&$	23.84 	$&$	23.84 	$&$	24.20 	$&$	24.48 	$&$	24.67 	$&$	24.43 	$&$	23.07 	$&$	20.61 	$\\
			$^{231}$Pa$\to^{207}$Tl$+^{24}$Ne$$&$	60.41 	$&$	1	$&$	22.89 	$&$	22.21 	$&$	22.20 	$&$	22.58 	$&$	22.88 	$&$	23.06 	$&$	22.82 	$&$	21.49 	$&$	19.07 	$\\
			$^{232}$U$\to^{208}$Pb$+^{24}$Ne$$&$	62.31 	$&$	0	$&$	20.39 	$&$	19.72 	$&$	19.72 	$&$	20.11 	$&$	20.42 	$&$	20.60 	$&$	20.35 	$&$	19.02 	$&$	16.63 	$\\
			$^{233}$U$\to^{209}$Pb$+^{24}$Ne$$&$	60.49 	$&$	2	$&$	24.84 	$&$	23.26 	$&$	23.26 	$&$	23.63 	$&$	23.92 	$&$	24.10 	$&$	23.86 	$&$	22.47 	$&$	20.04 	$\\
			$^{234}$U$\to^{210}$Pb$+^{24}$Ne$$&$	58.83 	$&$	0	$&$	25.93 	$&$	24.59 	$&$	24.59 	$&$	24.95 	$&$	25.23 	$&$	25.42 	$&$	25.18 	$&$	23.74 	$&$	21.28 	$\\
			$^{235}$U$\to^{211}$Pb$+^{24}$Ne$$&$	57.36 	$&$	1	$&$	27.42 	$&$	27.81 	$&$	27.81 	$&$	28.16 	$&$	28.43 	$&$	28.61 	$&$	28.37 	$&$	26.91 	$&$	24.42 	$\\
			$^{233}$U$\to^{208}$Pb$+^{25}$Ne$$&$	60.70 	$&$	2	$&$	24.84 	$&$	23.78 	$&$	23.78 	$&$	24.18 	$&$	24.48 	$&$	24.68 	$&$	24.42 	$&$	23.12 	$&$	20.72 	$\\
			$^{234}$U$\to^{208}$Pb$+^{26}$Ne$$&$	59.41 	$&$	0	$&$	25.93 	$&$	25.36 	$&$	25.36 	$&$	25.76 	$&$	26.07 	$&$	26.27 	$&$	26.01 	$&$	24.77 	$&$	22.36 	$\\
			$^{234}$U$\to^{206}$Hg$+^{28}$Mg$$&$	74.11 	$&$	0	$&$	25.53 	$&$	24.15 	$&$	24.15 	$&$	24.55 	$&$	24.87 	$&$	25.07 	$&$	24.81 	$&$	22.94 	$&$	20.53 	$\\
			$^{236}$U$\to^{208}$Hg$+^{28}$Mg$$&$	70.73 	$&$	0	$&$	27.58 	$&$	28.57 	$&$	28.57 	$&$	28.95 	$&$	29.23 	$&$	29.44 	$&$	29.18 	$&$	27.24 	$&$	24.77 	$\\
			$^{236}$Pu$\to^{208}$Pb$+^{28}$Mg$$&$	79.67 	$&$	0	$&$	21.52 	$&$	19.88 	$&$	19.88 	$&$	20.32 	$&$	20.68 	$&$	20.87 	$&$	20.60 	$&$	18.77 	$&$	16.44 	$\\
			$^{238}$Pu$\to^{210}$Pb$+^{28}$Mg$$&$	75.91 	$&$	0	$&$	25.70 	$&$	24.35 	$&$	24.34 	$&$	24.75 	$&$	25.07 	$&$	25.27 	$&$	25.00 	$&$	23.07 	$&$	20.67 	$\\
			$^{236}$U$\to^{206}$Hg$+^{30}$Mg$$&$	72.27 	$&$	0	$&$	27.58 	$&$	27.93 	$&$	27.93 	$&$	28.35 	$&$	28.67 	$&$	28.90 	$&$	28.62 	$&$	26.88 	$&$	24.49 	$\\
			$^{238}$Pu$\to^{208}$Pb$+^{30}$Mg$$&$	76.79 	$&$	0	$&$	25.70 	$&$	24.62 	$&$	24.62 	$&$	25.07 	$&$	25.42 	$&$	25.63 	$&$	25.35 	$&$	23.59 	$&$	21.26 	$\\
			$^{238}$Pu$\to^{206}$Hg$+^{32}$Si$$&$	91.19 	$&$	0	$&$	25.28 	$&$	24.17 	$&$	24.17 	$&$	24.62 	$&$	24.98 	$&$	25.20 	$&$	24.91 	$&$	22.56 	$&$	20.22 	$\\
			$^{242}$Cm$\to^{208}$Pb$+^{34}$Si$$&$	96.54 	$&$	0	$&$	23.15 	$&$	22.05 	$&$	22.05 	$&$	22.57 	$&$	22.98 	$&$	23.22 	$&$	22.90 	$&$	20.74 	$&$	18.52 	$\\
			\hline
			{Cluster}&$Q_{c}$&\ $\ell$ \ &&&&&$\rm{log}_{10}{\emph{T}}_{1/2}(s)$&&&\\
			\cline{4-12}
			decay&(MeV)&&\ EXP&Prox.2010&Dutt2011&Bass73&Bass77&Bass80&CW76&BW91&AW95 \\
			\hline
			$^{221}$Fr$\to^{207}$Tl$+^{14}$C$$&$	31.29 	$&$	3	$&$	14.56 	$&$	12.23 	$&$	12.07 	$&$	16.22 	$&$	15.77 	$&$	14.86 	$&$	13.38 	$&$	13.68 	$&$	13.64 	$\\
			$^{221}$Ra$\to^{207}$Pb$+^{14}$C$$&$	32.40 	$&$	3	$&$	13.39 	$&$	11.09 	$&$	10.97 	$&$	15.08 	$&$	14.63 	$&$	13.72 	$&$	12.28 	$&$	12.55 	$&$	12.51 	$\\
			$^{222}$Ra$\to^{208}$Pb$+^{14}$C$$&$	33.05 	$&$	0	$&$	11.22 	$&$	9.28 	$&$	9.15 	$&$	13.12 	$&$	12.61 	$&$	11.72 	$&$	10.45 	$&$	10.63 	$&$	10.62 	$\\
			$^{223}$Ra$\to^{209}$Pb$+^{14}$C$$&$	31.83 	$&$	4	$&$	15.05 	$&$	12.11 	$&$	11.96 	$&$	16.16 	$&$	15.71 	$&$	14.78 	$&$	13.26 	$&$	13.58 	$&$	13.54 	$\\
			$^{224}$Ra$\to^{210}$Pb$+^{14}$C$$&$	30.53 	$&$	0	$&$	15.87 	$&$	13.72 	$&$	13.55 	$&$	17.89 	$&$	17.46 	$&$	16.51 	$&$	14.86 	$&$	15.26 	$&$	15.20 	$\\
			$^{226}$Ra$\to^{212}$Pb$+^{14}$C$$&$	28.20 	$&$	0	$&$	21.20 	$&$	18.48 	$&$	18.28 	$&$	22.93 	$&$	22.55 	$&$	21.56 	$&$	19.60 	$&$	20.17 	$&$	20.08 	$\\
			$^{223}$Ac$\to^{209}$Bi$+^{14}$C$$&$	33.06 	$&$	2	$&$	12.60 	$&$	10.63 	$&$	10.53 	$&$	14.64 	$&$	14.17 	$&$	13.25 	$&$	11.83 	$&$	12.09 	$&$	12.06 	$\\
			$^{225}$Ac$\to^{211}$Bi$+^{14}$C$$&$	30.48 	$&$	4	$&$	17.16 	$&$	15.41 	$&$	15.27 	$&$	19.78 	$&$	19.39 	$&$	18.41 	$&$	16.59 	$&$	17.07 	$&$	17.00 	$\\
			$^{228}$Th$\to^{208}$Pb$+^{20}$O$$&$	44.72 	$&$	0	$&$	20.73 	$&$	18.96 	$&$	18.48 	$&$	24.14 	$&$	22.87 	$&$	21.87 	$&$	19.44 	$&$	20.12 	$&$	20.34 	$\\
			$^{231}$Pa$\to^{208}$Pb$+^{23}$F$$&$	51.88 	$&$	1	$&$	26.02 	$&$	22.40 	$&$	21.77 	$&$	28.07 	$&$	26.41 	$&$	25.34 	$&$	22.54 	$&$	23.35 	$&$	23.72 	$\\
			$^{230}$Th$\to^{206}$Hg$+^{24}$Ne$$&$	57.76 	$&$	0	$&$	24.63 	$&$	21.47 	$&$	20.89 	$&$	27.62 	$&$	25.91 	$&$	24.79 	$&$	21.60 	$&$	22.53 	$&$	22.92 	$\\
			$^{231}$Pa$\to^{207}$Tl$+^{24}$Ne$$&$	60.41 	$&$	1	$&$	22.89 	$&$	19.95 	$&$	19.41 	$&$	25.94 	$&$	24.19 	$&$	23.08 	$&$	20.11 	$&$	20.91 	$&$	21.34 	$\\
			$^{232}$U$\to^{208}$Pb$+^{24}$Ne$$&$	62.31 	$&$	0	$&$	20.39 	$&$	17.52 	$&$	17.03 	$&$	23.45 	$&$	21.67 	$&$	20.58 	$&$	17.71 	$&$	18.44 	$&$	18.89 	$\\
			$^{233}$U$\to^{209}$Pb$+^{24}$Ne$$&$	60.49 	$&$	2	$&$	24.84 	$&$	20.91 	$&$	20.39 	$&$	27.09 	$&$	25.35 	$&$	24.21 	$&$	21.07 	$&$	21.96 	$&$	22.38 	$\\
			$^{234}$U$\to^{210}$Pb$+^{24}$Ne$$&$	58.83 	$&$	0	$&$	25.93 	$&$	22.14 	$&$	21.58 	$&$	28.51 	$&$	26.79 	$&$	25.62 	$&$	22.26 	$&$	23.27 	$&$	23.68 	$\\
			$^{235}$U$\to^{211}$Pb$+^{24}$Ne$$&$	57.36 	$&$	1	$&$	27.42 	$&$	25.26 	$&$	24.68 	$&$	31.80 	$&$	30.10 	$&$	28.90 	$&$	25.36 	$&$	26.47 	$&$	26.86 	$\\
			$^{233}$U$\to^{208}$Pb$+^{25}$Ne$$&$	60.70 	$&$	2	$&$	24.84 	$&$	21.63 	$&$	21.01 	$&$	27.68 	$&$	25.77 	$&$	24.66 	$&$	21.63 	$&$	22.48 	$&$	22.95 	$\\
			$^{234}$U$\to^{208}$Pb$+^{26}$Ne$$&$	59.41 	$&$	0	$&$	25.93 	$&$	23.30 	$&$	22.54 	$&$	29.41 	$&$	27.35 	$&$	26.23 	$&$	23.11 	$&$	24.04 	$&$	24.54 	$\\
			$^{234}$U$\to^{206}$Hg$+^{28}$Mg$$&$	74.11 	$&$	0	$&$	25.53 	$&$	21.67 	$&$	21.08 	$&$	28.65 	$&$	26.45 	$&$	25.22 	$&$	21.50 	$&$	22.55 	$&$	23.16 	$\\
			$^{236}$U$\to^{208}$Hg$+^{28}$Mg$$&$	70.73 	$&$	0	$&$	27.58 	$&$	25.87 	$&$	25.21 	$&$	33.25 	$&$	31.09 	$&$	29.78 	$&$	25.63 	$&$	26.93 	$&$	27.50 	$\\
			$^{236}$Pu$\to^{208}$Pb$+^{28}$Mg$$&$	79.67 	$&$	0	$&$	21.52 	$&$	17.63 	$&$	17.15 	$&$	24.31 	$&$	22.04 	$&$	20.83 	$&$	17.54 	$&$	18.34 	$&$	19.00 	$\\
			$^{238}$Pu$\to^{210}$Pb$+^{28}$Mg$$&$	75.91 	$&$	0	$&$	25.70 	$&$	21.81 	$&$	21.26 	$&$	28.98 	$&$	26.76 	$&$	25.47 	$&$	21.64 	$&$	22.75 	$&$	23.37 	$\\
			$^{236}$U$\to^{206}$Hg$+^{30}$Mg$$&$	72.27 	$&$	0	$&$	27.58 	$&$	25.69 	$&$	24.84 	$&$	32.69 	$&$	30.21 	$&$	28.95 	$&$	25.14 	$&$	26.31 	$&$	26.98 	$\\
			$^{238}$Pu$\to^{208}$Pb$+^{30}$Mg$$&$	76.79 	$&$	0	$&$	25.70 	$&$	22.50 	$&$	21.76 	$&$	29.37 	$&$	26.83 	$&$	25.58 	$&$	22.04 	$&$	23.04 	$&$	23.75 	$\\
			$^{238}$Pu$\to^{206}$Hg$+^{32}$Si$$&$	91.19 	$&$	0	$&$	25.28 	$&$	21.67 	$&$	21.11 	$&$	29.34 	$&$	26.65 	$&$	25.30 	$&$	21.23 	$&$	22.33 	$&$	23.15 	$\\
			$^{242}$Cm$\to^{208}$Pb$+^{34}$Si$$&$	96.54 	$&$	0	$&$	23.15 	$&$	20.14 	$&$	19.45 	$&$	27.36 	$&$	24.26 	$&$	22.93 	$&$	19.45 	$&$	20.26 	$&$	21.23 	$\\
			\hline
			{Cluster}&$Q_{c}$&\ $\ell$ \ &&&&&$\rm{log}_{10}{\emph{T}}_{1/2}(s)$&&&\\
			\cline{4-12}
			decay&(MeV)&&\ EXP&Ng$\hat{o}$80&Denisov&Denisov DP&Guo2013&UDL&Ni&SL& \\
			\hline
			$^{221}$Fr$\to^{207}$Tl$+^{14}$C$$&$	31.29 	$&$	3	$&$	14.56 	$&$	15.06 	$&$	15.19 	$&$	8.11 	$&$	12.47 	$&$	12.70 	$&$	14.63 	$&$	13.54 	$&$	$\\
			$^{221}$Ra$\to^{207}$Pb$+^{14}$C$$&$	32.40 	$&$	3	$&$	13.39 	$&$	13.95 	$&$	14.01 	$&$	7.01 	$&$	11.36 	$&$	11.46 	$&$	13.48 	$&$	12.27 	$&$	$\\
			$^{222}$Ra$\to^{208}$Pb$+^{14}$C$$&$	33.05 	$&$	0	$&$	11.22 	$&$	12.05 	$&$	12.01 	$&$	5.30 	$&$	9.52 	$&$	10.07 	$&$	11.02 	$&$	11.00 	$&$	$\\
			$^{223}$Ra$\to^{209}$Pb$+^{14}$C$$&$	31.83 	$&$	4	$&$	15.05 	$&$	14.97 	$&$	15.10 	$&$	7.97 	$&$	12.35 	$&$	12.57 	$&$	14.56 	$&$	13.42 	$&$	$\\
			$^{224}$Ra$\to^{210}$Pb$+^{14}$C$$&$	30.53 	$&$	0	$&$	15.87 	$&$	16.62 	$&$	16.85 	$&$	9.49 	$&$	13.97 	$&$	15.38 	$&$	15.86 	$&$	16.14 	$&$	$\\
			$^{226}$Ra$\to^{212}$Pb$+^{14}$C$$&$	28.20 	$&$	0	$&$	21.20 	$&$	21.50 	$&$	21.94 	$&$	14.06 	$&$	18.77 	$&$	20.95 	$&$	20.94 	$&$	21.53 	$&$	$\\
			$^{223}$Ac$\to^{209}$Bi$+^{14}$C$$&$	33.06 	$&$	2	$&$	12.60 	$&$	13.50 	$&$	13.53 	$&$	6.58 	$&$	10.90 	$&$	11.08 	$&$	13.19 	$&$	11.93 	$&$	$\\
			$^{225}$Ac$\to^{211}$Bi$+^{14}$C$$&$	30.48 	$&$	4	$&$	17.16 	$&$	18.43 	$&$	18.75 	$&$	11.09 	$&$	15.71 	$&$	16.61 	$&$	18.24 	$&$	17.26 	$&$	$\\
			$^{228}$Th$\to^{208}$Pb$+^{20}$O$$&$	44.72 	$&$	0	$&$	20.73 	$&$	22.09 	$&$	22.98 	$&$	13.36 	$&$	18.73 	$&$	21.97 	$&$	21.54 	$&$	21.20 	$&$	$\\
			$^{231}$Pa$\to^{208}$Pb$+^{23}$F$$&$	51.88 	$&$	1	$&$	26.02 	$&$	25.62 	$&$	26.75 	$&$	16.31 	$&$	21.92 	$&$	24.90 	$&$	25.59 	$&$	23.78 	$&$	$\\
			$^{230}$Th$\to^{206}$Hg$+^{24}$Ne$$&$	57.76 	$&$	0	$&$	24.63 	$&$	24.95 	$&$	26.40 	$&$	15.17 	$&$	21.07 	$&$	25.39 	$&$	24.58 	$&$	23.92 	$&$	$\\
			$^{231}$Pa$\to^{207}$Tl$+^{24}$Ne$$&$	60.41 	$&$	1	$&$	22.89 	$&$	23.36 	$&$	24.69 	$&$	13.64 	$&$	19.51 	$&$	22.27 	$&$	23.09 	$&$	21.32 	$&$	$\\
			$^{232}$U$\to^{208}$Pb$+^{24}$Ne$$&$	62.31 	$&$	0	$&$	20.39 	$&$	20.92 	$&$	22.17 	$&$	11.21 	$&$	17.07 	$&$	20.59 	$&$	20.36 	$&$	19.94 	$&$	$\\
			$^{233}$U$\to^{209}$Pb$+^{24}$Ne$$&$	60.49 	$&$	2	$&$	24.84 	$&$	24.41 	$&$	25.80 	$&$	14.57 	$&$	20.51 	$&$	23.63 	$&$	24.41 	$&$	22.55 	$&$	$\\
			$^{234}$U$\to^{210}$Pb$+^{24}$Ne$$&$	58.83 	$&$	0	$&$	25.93 	$&$	25.71 	$&$	27.22 	$&$	15.80 	$&$	21.78 	$&$	26.52 	$&$	25.81 	$&$	25.03 	$&$	$\\
			$^{235}$U$\to^{211}$Pb$+^{24}$Ne$$&$	57.36 	$&$	1	$&$	27.42 	$&$	28.90 	$&$	30.51 	$&$	18.96 	$&$	24.95 	$&$	29.16 	$&$	29.51 	$&$	27.31 	$&$	$\\
			$^{233}$U$\to^{208}$Pb$+^{25}$Ne$$&$	60.70 	$&$	2	$&$	24.84 	$&$	25.00 	$&$	26.29 	$&$	15.22 	$&$	21.06 	$&$	24.00 	$&$	24.88 	$&$	23.05 	$&$	$\\
			$^{234}$U$\to^{208}$Pb$+^{26}$Ne$$&$	59.41 	$&$	0	$&$	25.93 	$&$	26.60 	$&$	27.91 	$&$	16.83 	$&$	22.58 	$&$	27.01 	$&$	26.52 	$&$	25.84 	$&$	$\\
			$^{234}$U$\to^{206}$Hg$+^{28}$Mg$$&$	74.11 	$&$	0	$&$	25.53 	$&$	25.45 	$&$	27.22 	$&$	14.76 	$&$	21.10 	$&$	25.77 	$&$	25.25 	$&$	24.76 	$&$	$\\
			$^{236}$U$\to^{208}$Hg$+^{28}$Mg$$&$	70.73 	$&$	0	$&$	27.58 	$&$	29.80 	$&$	31.80 	$&$	19.14 	$&$	25.43 	$&$	31.25 	$&$	30.33 	$&$	29.28 	$&$	$\\
			$^{236}$Pu$\to^{208}$Pb$+^{28}$Mg$$&$	79.67 	$&$	0	$&$	21.52 	$&$	21.30 	$&$	22.83 	$&$	10.65 	$&$	16.96 	$&$	20.64 	$&$	20.76 	$&$	20.83 	$&$	$\\
			$^{238}$Pu$\to^{210}$Pb$+^{28}$Mg$$&$	75.91 	$&$	0	$&$	25.70 	$&$	25.67 	$&$	27.48 	$&$	14.87 	$&$	21.27 	$&$	26.26 	$&$	25.96 	$&$	25.42 	$&$	$\\
			$^{236}$U$\to^{206}$Hg$+^{30}$Mg$$&$	72.27 	$&$	0	$&$	27.58 	$&$	29.30 	$&$	31.04 	$&$	18.79 	$&$	24.82 	$&$	29.94 	$&$	29.47 	$&$	28.69 	$&$	$\\
			$^{238}$Pu$\to^{208}$Pb$+^{30}$Mg$$&$	76.79 	$&$	0	$&$	25.70 	$&$	26.08 	$&$	27.66 	$&$	15.42 	$&$	21.58 	$&$	26.06 	$&$	26.10 	$&$	25.71 	$&$	$\\
			$^{238}$Pu$\to^{206}$Hg$+^{32}$Si$$&$	91.19 	$&$	0	$&$	25.28 	$&$	25.70 	$&$	27.66 	$&$	14.16 	$&$	20.89 	$&$	25.48 	$&$	25.59 	$&$	25.70 	$&$	$\\
			$^{242}$Cm$\to^{208}$Pb$+^{34}$Si$$&$	96.54 	$&$	0	$&$	23.15 	$&$	23.79 	$&$	25.41 	$&$	12.38 	$&$	18.90 	$&$	22.35 	$&$	23.48 	$&$	24.21 	$&$	$\\
			\hline
			\hline
	\end{longtable*}
\endgroup

\begin{figure}[h]\centering
	\includegraphics[width=9cm]{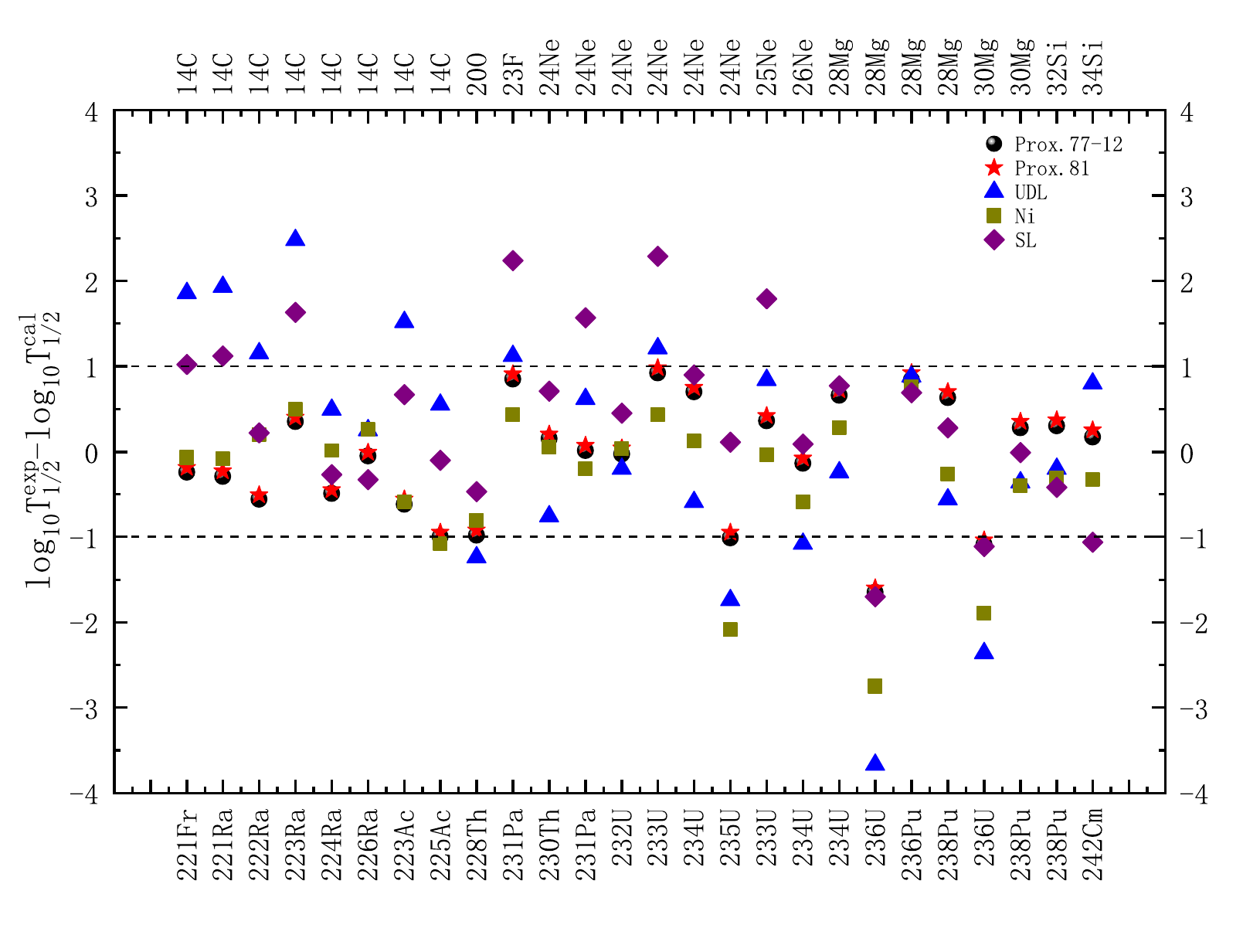}
	\caption{(color online) Comparison of the discrepancy between the experimental cluster radioactivity half-lives and the calculated ones using CPPM with Prox.77-12 and Prox.81 as well as the compared formulae in logarithmic form.}
	\label {fig 2}
\end{figure}

In order to intuitively survey the deviations between the calculated cluster radioactivity half-lives and experimental ones, we adopt the root-mean-square deviation $\sigma$ as a measure in this work. It can be expressed as
\begin{eqnarray}\label{47}
\sigma = \sqrt{\sum\limits_{i=1}^{n}\frac{({\rm log}_{10}{T_{1/2}^{{exp,i}}}-{\rm log}_{10}{T_{1/2}^{{cal,i}}})^2}{n}},
\end{eqnarray}
where ${\rm log}_{10}{T_{1/2}^{{exp,i}}}$ and ${\rm log}_{10}{T_{1/2}^{{cal,i}}}$ represent the logarithmic forms of the experimental cluster radioactivity half-lives and calculated ones for the $i$-th nucleus, respectively. $n$ is the number of nuclei involved for different decay cases. The detailed calculations of root-mean-square deviation $\sigma$ for 28 different versions of proximity potential formalisms are listed in Table \ref{Tab3}. Additionally, the root-mean-square deviation $\sigma$ obtained by using UDL, Ni's empirical formula and SL are also presented in this table. From this table, the most suitable proximity potential formalisms for the cluster radioactivity can be obtained in Prox.77-12 and Prox.81 since both of them have the lowest root-mean-square deviation $\sigma=0.681$. Simultaneously, some other proximity potential formalisms with the root-mean-square deviation $\sigma$ less than 1 can also be splendidly applied to the cluster radioactivity. There are also some proximity potential calculations with huge deviations from the experimental data such as Denisov DP, Prox.00 DP, Guo2013, Dutt2011 and so on. On the whole, the calculations by CPPM with different versions of proximity potential formalisms have a comparable accuracy with experimental data. In order to further verify the feasibility of applying proximity potential to the cluster radioactivity, we plot the differences between the experimental cluster radioactivity half-lives and the calculated ones by using CPPM with Prox.77-12 and Prox.81 as well as the compared formulae in logarithmic form in Fig. \ref{fig 2}. 
From this figure, we can clearly see that the deviations using CPPM with Prox.77-12 and Prox.81 are mainly among -1$\rightarrow$1, whereas the deviation distribution of UDL and SL are slightly dispersed. It demonstrates that Prox.77-12 and Prox.81 can be adopted to obtain the most precise calculations of cluster radioactivity half-lives among 28 versions of proximity potential formalisms.

\begingroup
\renewcommand*{\arraystretch}{1.3}
\setlength{\tabcolsep}{1mm}
\setlength\LTleft{0pt}
\setlength\LTright{0pt}
\setlength{\LTcapwidth}{7in}
\begin{longtable*}	{@{\extracolsep{\fill}} ccccccccc}
	\caption{The root-mean-square deviation $\sigma$ between the experimental data and the calculated ones by using CPPM with 28 different versions of the proximity potential formalisms, UDL, Ni's empirical formula and SL for cluster radioactivity.}
	\label {Tab3} \\
		\hline 
        \hline 
		\endfirsthead
		\multicolumn{9}{c}
		{{\tablename\ \thetable{} -- continued from previous page}} \\
		\hline
		\endhead
		\hline \multicolumn{9}{r}{{Continued on next page}}\\
		\endfoot
		\endlastfoot
	    {Method}&Prox.77-1&Prox.77-2&Prox.77-3&Prox.77-4&Prox.77-5&Prox.77-6&Prox.77-7&Prox.77-8 \\  	
		$\sigma$&0.686&0.691& 	1.100& 	0.838& 	0.815& 	0.943& 	0.914& 	0.700 \\
		\hline
		{Method}&Prox.77-9&Prox.77-10&Prox.77-11&Prox.77-12&Prox.77-13&Prox.81&Prox.00&Prox.00 DP \\  
		$\sigma$&0.924 &0.924 &	0.730 &	0.681 &	0.715 &	0.681 &	1.594 &	3.771 \\
		\hline
		{Method}&Prox.2010&Dutt2011&Bass73&Bass77&Bass80&CW76&BW91&AW95\\
		$\sigma$&2.915 &3.391 &	3.146 &	1.566 & 0.763 &	2.779 &	2.006 &	1.594 \\
		\hline
		{Method}&Ng$\hat{o}$80&Denisov&Denisov DP&Guo2013&UDL&Ni&SL&\\
		$\sigma$&0.844&1.878 &8.859 &3.301 &1.374&0.875 &1.072 & \\
		\hline 
		\hline 		
	\end{longtable*}
\endgroup

\begin{figure}[h]\centering
	\includegraphics[width=9cm]{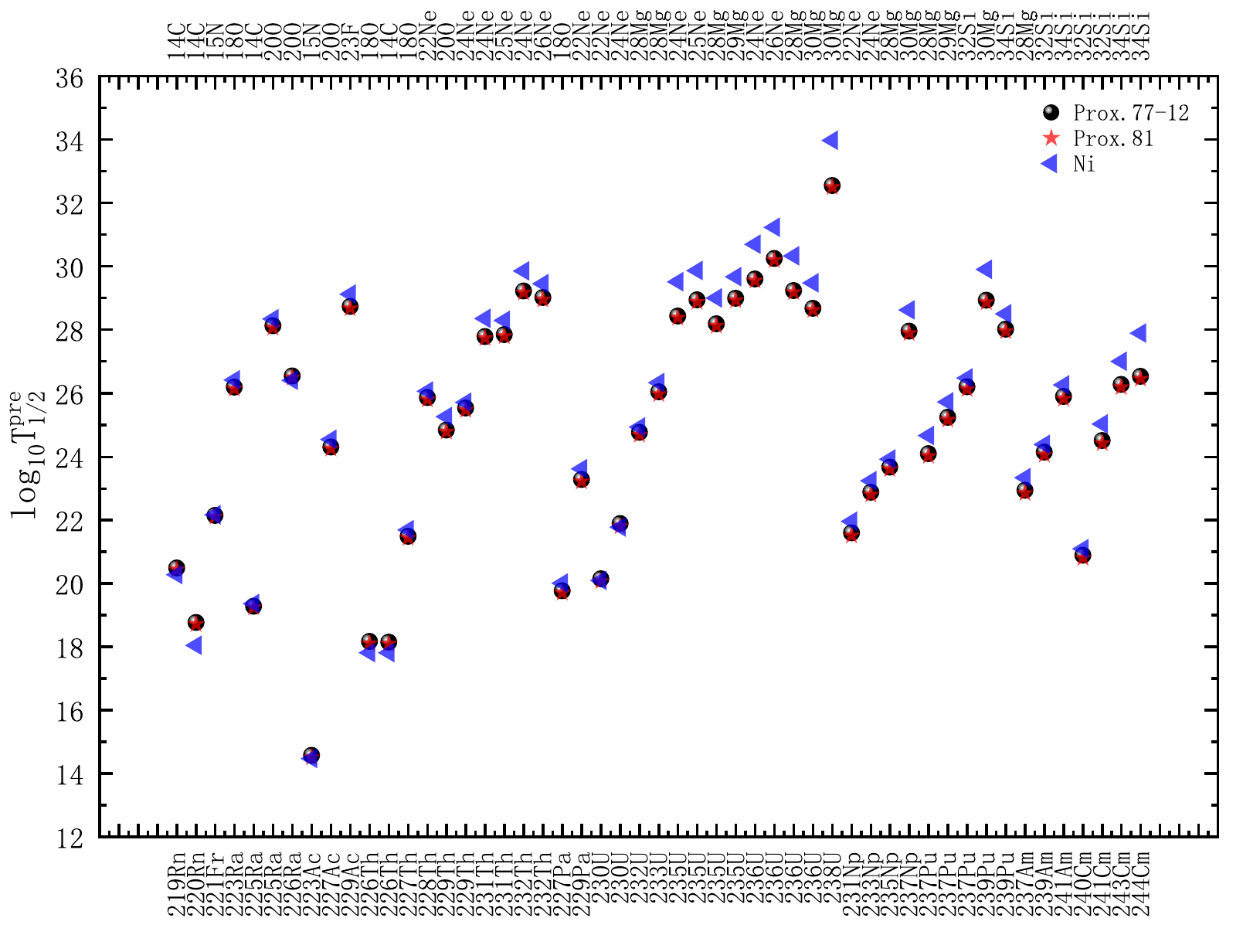}
	\caption{(color online) The predicted half-lives are obtained by using CPPM with Prox.77-12 and Prox.81 as well as Ni's empirical formula in logarithmic form.}
	\label {fig 3}
\end{figure}

Encouraged by the good agreement between the experimental cluster radioactivity half-lives and the calculated ones using CPPM with Prox.77-12 and Prox.81.
As an application, we employ four proximity potentials with the smallest root-mean-square deviation to predict the half-lives of 51 possible cluster radioactive candidates, whose cluster radioactivity is energetically allowed or observed but not yet quantified in NUBASE2020. Meanwhile, we also calculated their $\alpha$ decay half-lives compared with the cluster radioactivity ones in order to determine the most dominant decay modes of these predicted nuclei. All the predicted results are listed in Table \ref{Tab4}. 
In this table, the first to fourth columns are the parent nuclei, the corresponding emitted particles, the decay energies $Q$ of $\alpha$ decay and cluster radioactivity as well as the orbital angular momentum $\ell$ carried by the emitted particles, respectively. The last six columns represent the experimental half-lifes of $\alpha$ decay and the cluster radioactivity as well as the predicted results obtained by using CPPM with Prox.77-12, Prox.81, Prox.77-1 and Prox.77-2 as well as Ni's empirical formula in logarithmic form, respectively. The decay energies and experimental half-lifes of $\alpha$ decay are taken from AME2020 \cite{2021CPC45030003} and NUBASE2020 \cite{2021CPC45030001}. At the same time, we introduce the branching ratio of cluster radioactivity relative to the $\alpha$ decay $\varsigma=log_{10}{\emph{T}}_{1/2}^{\ \alpha}-log_{10}{\emph{T}}_{1/2}^{\ C}$ \cite{2012PRC85034615} to manifest the competition among the two. 
From this table, we can find that the rang of $\varsigma$ from about $-11$ to $-20$ is significantly less than 0, which indicates that these predicted nuclei are more likely to undergo $\alpha$ decay than cluster radioactivity. In addition, the predicted results by using CPPM with four proximity potentials and Ni's empirical formula remained roughly the same order of magnitude for cluster radioactivity and marvellously reproduce the experimental data for $\alpha$ decay. In order to intuitively investigate the agreement of our predicted results with Ni's empirical formula in cluster radioactivity, we plot their the logarithmic half-lives of CPPM with minimum root-mean-square deviation proximity potentials and Ni's empirical formula in Fig. \ref{fig 3}. From this figure, it is obviously that the calculated results by using CPPM with Prox.77-12 and Prox.81 are very close to Ni's empirical formula, which indicates that CPPM with proximity potential is reliable for calculating cluster radioactivity half-lives. Meanwhile, we also hope that these predicted results will be useful for exploring new cluster radioactive nuclei in the future experimental.

\begingroup
\renewcommand*{\arraystretch}{1.5}
\setlength{\tabcolsep}{1mm}
\setlength\LTleft{0pt}
\setlength\LTright{0pt}
\setlength{\LTcapwidth}{7in}
\begin{longtable*}
	{@{\extracolsep{\fill}} cccccccccc}
			\caption{Predicted half-lives for 51 possible cluster radioactive nuclei.}
			\label{Tab4}\\
			\hline 
			\hline
			Parent&Emitted&$Q$&$\ell$&&&&$\rm{log}_{10}{\emph{T}}_{1/2}(s)$&&\\
			\cline{5-10}
			nuclei&particles&(MeV)&&EXP&Prox.77-12&Prox.81&Prox.77-1&Prox.77-2&Ni\\
			\hline                                 
			\endfirsthead
			\multicolumn{10}{c}
			{{\tablename\ \thetable{} -- continued from previous page}} \\
			\hline 
			\endhead
			\hline \multicolumn{10}{r}{{Continued on next page}}\\
			\endfoot
			\endlastfoot
			
			$^{219}$Rn&	$^{4}$He&$	6.95 	$&$	2	$&$	0.60 	$&$	0.21 	$&$	0.18 	$&$	0.22 	$&$	0.18 	$&$	0.55 	$\\
			$^{}$&	$^{14}$C&$	28.10 	$&$	3 	$&$	-	$&$	20.48 	$&$	20.43 	$&$	20.50 	$&$	20.39 	$&$	20.28 	$\\
			$^{220}$Rn&	$^{4}$He&$	6.40 	$&$	0	$&$	1.75 	$&$	2.00 	$&$	1.97 	$&$	2.01 	$&$	1.96 	$&$	1.93 	$\\
			$^{}$&	$^{14}$C&$	28.54 	$&$	0 	$&$	-	$&$	18.76 	$&$	18.71 	$&$	18.78 	$&$	18.67 	$&$	18.05 	$\\
			$^{221}$Fr&	$^{4}$He&$	6.46 	$&$	2	$&$	2.46 	$&$	2.55 	$&$	2.52 	$&$	2.56 	$&$	2.51 	$&$	2.71 	$\\
			$^{}$&	$^{15}$N&$	34.12 	$&$	3 	$&$	-	$&$	22.14 	$&$	22.10 	$&$	22.16 	$&$	22.06 	$&$	22.16 	$\\
			$^{223}$Ra&	$^{4}$He&$	5.98 	$&$	2	$&$	5.99 	$&$	5.14 	$&$	5.11 	$&$	5.14 	$&$	5.10 	$&$	5.39 	$\\
			$^{}$&	$^{18}$O&$	40.30 	$&$	1 	$&$	-	$&$	26.19 	$&$	26.14 	$&$	26.21 	$&$	26.09 	$&$	26.42 	$\\
			$^{225}$Ra&	$^{4}$He&$	5.10 	$&$	4 	$&$	-	$&$	10.43 	$&$	10.40 	$&$	10.44 	$&$	10.40 	$&$	10.01 	$\\
			$^{}$&	$^{14}$C&$	29.47 	$&$	4 	$&$	-	$&$	19.28 	$&$	19.24 	$&$	19.31 	$&$	19.20 	$&$	19.36 	$\\
			$^{}$&	$^{20}$O&$	40.48 	$&$	1 	$&$	-	$&$	28.12 	$&$	28.07 	$&$	28.15 	$&$	28.01 	$&$	28.35 	$\\
			$^{226}$Ra&	$^{4}$He&$	4.87 	$&$	0	$&$	10.70 	$&$	10.96 	$&$	10.93 	$&$	10.97 	$&$	10.93 	$&$	10.67 	$\\
			$^{}$&	$^{20}$O&$	40.82 	$&$	0 	$&$	-	$&$	26.54 	$&$	26.49 	$&$	26.57 	$&$	26.44 	$&$	26.41 	$\\
			$^{223}$Ac&	$^{4}$He&$	6.78 	$&$	2	$&$	2.10 	$&$	2.10 	$&$	2.07 	$&$	2.10 	$&$	2.06 	$&$	2.27 	$\\
			$^{}$&	$^{15}$N&$	39.47 	$&$	3 	$&$	>14.76	$&$	14.57 	$&$	14.51 	$&$	14.58 	$&$	14.47 	$&$	14.47 	$\\
			$^{227}$Ac&	$^{4}$He&$	5.04 	$&$	0	$&$	10.70 	$&$	10.53 	$&$	10.50 	$&$	10.53 	$&$	10.50 	$&$	10.73 	$\\
			$^{}$&	$^{20}$O&$	43.09 	$&$	1 	$&$	-	$&$	24.30 	$&$	24.25 	$&$	24.33 	$&$	24.19 	$&$	24.54 	$\\
			$^{229}$Ac&	$^{4}$He&$	4.44 	$&$	1 	$&$	-	$&$	14.76 	$&$	14.74 	$&$	14.77 	$&$	14.73 	$&$	14.75 	$\\
			$^{}$&	$^{23}$F&$	48.35 	$&$	2 	$&$	-	$&$	28.73 	$&$	28.68 	$&$	28.77 	$&$	28.61 	$&$	29.13 	$\\
			$^{226}$Th&	$^{4}$He&$	6.45 	$&$	0	$&$	3.27 	$&$	3.51 	$&$	3.48 	$&$	3.52 	$&$	3.48 	$&$	3.44 	$\\
			$^{}$&	$^{18}$O&$	45.73 	$&$	0 	$&$	-	$&$	18.17 	$&$	18.11 	$&$	18.19 	$&$	18.06 	$&$	17.81 	$\\
			$^{}$&	$^{14}$C&$	30.55 	$&$	0 	$&$	>16.76	$&$	18.14 	$&$	18.09 	$&$	18.16 	$&$	18.05 	$&$	17.82 	$\\
			$^{227}$Th&	$^{4}$He&$	6.15 	$&$	2	$&$	6.21 	$&$	5.21 	$&$	5.18 	$&$	5.22 	$&$	5.18 	$&$	5.51 	$\\
			$^{}$&	$^{18}$O&$	44.20 	$&$	4 	$&$	>15.36	$&$	21.48 	$&$	21.43 	$&$	21.50 	$&$	21.37 	$&$	21.69 	$\\
			$^{228}$Th&	$^{4}$He&$	5.52 	$&$	0	$&$	7.78 	$&$	8.06 	$&$	8.03 	$&$	8.06 	$&$	8.02 	$&$	7.88 	$\\
			$^{}$&	$^{22}$Ne&$	55.74 	$&$	0 	$&$	-	$&$	25.86 	$&$	25.80 	$&$	25.88 	$&$	25.74 	$&$	26.07 	$\\
			$^{229}$Th&	$^{4}$He&$	5.17 	$&$	2	$&$	11.40 	$&$	10.45 	$&$	10.43 	$&$	10.46 	$&$	10.42 	$&$	10.61 	$\\
			$^{}$&	$^{20}$O&$	43.40 	$&$	2 	$&$	-	$&$	24.84 	$&$	24.79 	$&$	24.87 	$&$	24.73 	$&$	25.26 	$\\
			$^{}$&	$^{24}$Ne&$	57.83 	$&$	3 	$&$	-	$&$	25.53 	$&$	25.47 	$&$	25.56 	$&$	25.40 	$&$	25.71 	$\\
			$^{231}$Th&	$^{4}$He&$	4.21 	$&$	2 	$&$	-	$&$	17.35 	$&$	17.33 	$&$	17.36 	$&$	17.32 	$&$	17.24 	$\\
			$^{}$&	$^{24}$Ne&$	56.25 	$&$	2 	$&$	-	$&$	27.78 	$&$	27.73 	$&$	27.82 	$&$	27.66 	$&$	28.36 	$\\
			$^{}$&	$^{25}$Ne&$	56.80 	$&$	2 	$&$	-	$&$	27.85 	$&$	27.80 	$&$	27.89 	$&$	27.73 	$&$	28.30 	$\\
			$^{232}$Th&	$^{4}$He&$	4.08 	$&$	0	$&$	17.65 	$&$	18.09 	$&$	18.07 	$&$	18.10 	$&$	18.06 	$&$	17.57 	$\\
			$^{}$&	$^{24}$Ne&$	54.67 	$&$	0 	$&$	>29.20	$&$	29.22 	$&$	29.17 	$&$	29.26 	$&$	29.11 	$&$	29.86 	$\\
			$^{}$&	$^{26}$Ne&$	55.91 	$&$	0 	$&$	>29.20	$&$	29.01 	$&$	28.96 	$&$	29.06 	$&$	28.89 	$&$	29.45 	$\\
			$^{227}$Pa&	$^{4}$He&$	6.58 	$&$	0	$&$	3.43 	$&$	3.52 	$&$	3.49 	$&$	3.52 	$&$	3.48 	$&$	3.92 	$\\
			$^{}$&	$^{18}$O&$	45.87 	$&$	2 	$&$	-	$&$	19.77 	$&$	19.70 	$&$	19.78 	$&$	19.65 	$&$	20.01 	$\\
			$^{229}$Pa&	$^{4}$He&$	5.84 	$&$	1	$&$	7.43 	$&$	7.04 	$&$	7.01 	$&$	7.04 	$&$	7.00 	$&$	7.30 	$\\
			$^{}$&	$^{22}$Ne&$	58.96 	$&$	2 	$&$	-	$&$	23.27 	$&$	23.21 	$&$	23.29 	$&$	23.15 	$&$	23.62 	$\\
			$^{230}$U&	$^{4}$He&$	5.99 	$&$	0	$&$	6.24 	$&$	6.54 	$&$	6.51 	$&$	6.54 	$&$	6.50 	$&$	6.41 	$\\
			$^{}$&	$^{22}$Ne&$	61.39 	$&$	0 	$&$	>18.20	$&$	20.14 	$&$	20.07 	$&$	20.15 	$&$	20.00 	$&$	20.09 	$\\
			$^{}$&	$^{24}$Ne&$	61.35 	$&$	0 	$&$	>18.20	$&$	21.88 	$&$	21.81 	$&$	21.90 	$&$	21.73 	$&$	21.78 	$\\
			$^{232}$U&	$^{4}$He&$	5.41 	$&$	0	$&$	9.34 	$&$	9.65 	$&$	9.62 	$&$	9.66 	$&$	9.62 	$&$	9.46 	$\\
			$^{}$&	$^{28}$Mg&$	74.32 	$&$	0 	$&$	>22.26	$&$	24.76 	$&$	24.69 	$&$	24.79 	$&$	24.61 	$&$	24.93 	$\\
			$^{233}$U&	$^{4}$He&$	4.91 	$&$	0	$&$	12.70 	$&$	12.94 	$&$	12.91 	$&$	12.94 	$&$	12.90 	$&$	13.25 	$\\
			$^{}$&	$^{28}$Mg&$	74.23 	$&$	3 	$&$	>27.59	$&$	26.04 	$&$	25.97 	$&$	26.07 	$&$	25.89 	$&$	26.33 	$\\
			$^{235}$U&	$^{4}$He&$	4.68 	$&$	1 	$&$	16.35 	$&$	14.61 	$&$	14.58 	$&$	14.61 	$&$	14.58 	$&$	14.82 	$\\
			$^{}$&	$^{24}$Ne&$	57.36 	$&$	1 	$&$	>27.65	$&$	28.43 	$&$	28.38 	$&$	28.46 	$&$	28.31 	$&$	29.51 	$\\
			$^{}$&	$^{25}$Ne&$	57.68 	$&$	3 	$&$	>27.65	$&$	28.94 	$&$	28.88 	$&$	28.97 	$&$	28.81 	$&$	29.87 	$\\
			$^{}$&	$^{28}$Mg&$	72.43 	$&$	1 	$&$	>28.45	$&$	28.19 	$&$	28.13 	$&$	28.22 	$&$	28.05 	$&$	29.00 	$\\
			$^{}$&	$^{29}$Mg&$	72.48 	$&$	3 	$&$	>28.45	$&$	28.99 	$&$	28.93 	$&$	29.03 	$&$	28.85 	$&$	29.67 	$\\
			$^{236}$U&	$^{4}$He&$	4.57 	$&$	0	$&$	14.87 	$&$	15.20 	$&$	15.18 	$&$	15.21 	$&$	15.17 	$&$	14.86 	$\\
			$^{}$&	$^{24}$Ne&$	55.95 	$&$	0 	$&$	>26.27	$&$	29.60 	$&$	29.55 	$&$	29.64 	$&$	29.49 	$&$	30.70 	$\\
			$^{}$&	$^{26}$Ne&$	56.69 	$&$	0 	$&$	>26.27	$&$	30.25 	$&$	30.20 	$&$	30.29 	$&$	30.12 	$&$	31.23 	$\\
			$^{}$&	$^{28}$Mg&$	70.73 	$&$	0 	$&$	>26.27	$&$	29.24 	$&$	29.19 	$&$	29.28 	$&$	29.11 	$&$	30.33 	$\\
			$^{}$&	$^{30}$Mg&$	72.27 	$&$	0 	$&$	>26.27	$&$	28.67 	$&$	28.62 	$&$	28.72 	$&$	28.53 	$&$	29.48 	$\\
			$^{238}$U&	$^{4}$He&$	4.27 	$&$	0	$&$	17.15 	$&$	17.58 	$&$	17.56 	$&$	17.59 	$&$	17.55 	$&$	17.17 	$\\
			$^{}$&	$^{30}$Mg&$	69.46 	$&$	0 	$&$	-	$&$	32.55 	$&$	32.50 	$&$	32.60 	$&$	32.42 	$&$	33.98 	$\\
			$^{231}$Np&	$^{4}$He&$	6.37 	$&$	1	$&$	5.14 	$&$	5.38 	$&$	5.35 	$&$	5.38 	$&$	5.34 	$&$	5.69 	$\\
			$^{}$&	$^{22}$Ne&$	61.90 	$&$	3 	$&$	-	$&$	21.59 	$&$	21.51 	$&$	21.60 	$&$	21.45 	$&$	21.96 	$\\
			$^{233}$Np&	$^{4}$He&$	5.63 	$&$	0	$&$	8.49 	$&$	9.02 	$&$	8.99 	$&$	9.03 	$&$	8.99 	$&$	9.33 	$\\
			$^{}$&	$^{24}$Ne&$	62.16 	$&$	3 	$&$	-	$&$	22.87 	$&$	22.80 	$&$	22.89 	$&$	22.72 	$&$	23.24 	$\\
			$^{235}$Np&	$^{4}$He&$	5.19 	$&$	1	$&$	12.12 	$&$	11.68 	$&$	11.65 	$&$	11.69 	$&$	11.65 	$&$	11.85 	$\\
			$^{}$&	$^{28}$Mg&$	77.10 	$&$	2 	$&$	-	$&$	23.67 	$&$	23.60 	$&$	23.70 	$&$	23.51 	$&$	23.92 	$\\
			$^{237}$Np&	$^{4}$He&$	4.96 	$&$	1	$&$	13.83 	$&$	13.15 	$&$	13.13 	$&$	13.16 	$&$	13.12 	$&$	13.31 	$\\
			$^{}$&	$^{30}$Mg&$	74.79 	$&$	2 	$&$	>27.57	$&$	27.96 	$&$	27.90 	$&$	28.00 	$&$	27.81 	$&$	28.63 	$\\
			$^{237}$Pu&	$^{4}$He&$	5.75 	$&$	1 	$&$	6.60 	$&$	8.86 	$&$	8.84 	$&$	8.87 	$&$	8.83 	$&$	9.31 	$\\
			$^{}$&	$^{28}$Mg&$	77.73 	$&$	1 	$&$	-	$&$	24.09 	$&$	24.02 	$&$	24.12 	$&$	23.93 	$&$	24.66 	$\\
			$^{}$&	$^{29}$Mg&$	77.45 	$&$	3 	$&$	-	$&$	25.24 	$&$	25.17 	$&$	25.27 	$&$	25.08 	$&$	25.73 	$\\
			$^{}$&	$^{32}$Si&$	91.46 	$&$	4 	$&$	-	$&$	26.20 	$&$	26.12 	$&$	26.23 	$&$	26.03 	$&$	26.48 	$\\
			$^{239}$Pu&	$^{4}$He&$	5.24 	$&$	0	$&$	11.88 	$&$	11.75 	$&$	11.73 	$&$	11.76 	$&$	11.72 	$&$	12.20 	$\\
			$^{}$&	$^{30}$Mg&$	75.08 	$&$	4 	$&$	-	$&$	28.93 	$&$	28.87 	$&$	28.97 	$&$	28.78 	$&$	29.90 	$\\
			$^{}$&	$^{34}$Si&$	90.87 	$&$	1 	$&$	-	$&$	28.02 	$&$	27.95 	$&$	28.06 	$&$	27.85 	$&$	28.50 	$\\
			$^{237}$Am&	$^{4}$He&$	6.20 	$&$	1	$&$	7.24 	$&$	7.02 	$&$	6.99 	$&$	7.02 	$&$	6.98 	$&$	7.36 	$\\
			$^{}$&	$^{28}$Mg&$	79.85 	$&$	2 	$&$	-	$&$	22.93 	$&$	22.85 	$&$	22.95 	$&$	22.76 	$&$	23.34 	$\\
			$^{239}$Am&	$^{4}$He&$	5.92 	$&$	1	$&$	8.63 	$&$	8.41 	$&$	8.38 	$&$	8.42 	$&$	8.38 	$&$	8.74 	$\\
			$^{}$&	$^{32}$Si&$	94.50 	$&$	3 	$&$	-	$&$	24.14 	$&$	24.06 	$&$	24.17 	$&$	23.97 	$&$	24.38 	$\\
			$^{241}$Am&	$^{4}$He&$	5.64 	$&$	1	$&$	10.14 	$&$	9.91 	$&$	9.89 	$&$	9.92 	$&$	9.88 	$&$	10.23 	$\\
			$^{}$&	$^{34}$Si&$	93.96 	$&$	3 	$&$	>24.41	$&$	25.90 	$&$	25.82 	$&$	25.94 	$&$	25.72 	$&$	26.26 	$\\
			$^{240}$Cm&	$^{4}$He&$	6.40 	$&$	0	$&$	6.42 	$&$	6.27 	$&$	6.24 	$&$	6.27 	$&$	6.23 	$&$	6.26 	$\\
			$^{}$&	$^{32}$Si&$	97.55 	$&$	0 	$&$	-	$&$	20.89 	$&$	20.80 	$&$	20.92 	$&$	20.70 	$&$	21.09 	$\\
			$^{241}$Cm&	$^{4}$He&$	6.19 	$&$	3	$&$	8.45 	$&$	7.84 	$&$	7.81 	$&$	7.84 	$&$	7.81 	$&$	8.00 	$\\
			$^{}$&	$^{32}$Si&$	95.39 	$&$	4 	$&$	-	$&$	24.50 	$&$	24.41 	$&$	24.52 	$&$	24.32 	$&$	25.03 	$\\
			$^{243}$Cm&	$^{4}$He&$	6.17 	$&$	2	$&$	8.96 	$&$	7.68 	$&$	7.65 	$&$	7.69 	$&$	7.65 	$&$	8.10 	$\\
			$^{}$&	$^{34}$Si&$	94.79 	$&$	2 	$&$	-	$&$	26.27 	$&$	26.19 	$&$	26.31 	$&$	26.09 	$&$	27.00 	$\\
			$^{244}$Cm&	$^{4}$He&$	5.90 	$&$	0	$&$	8.76 	$&$	8.71 	$&$	8.69 	$&$	8.72 	$&$	8.68 	$&$	8.71 	$\\
			$^{}$&	$^{34}$Si&$	93.17 	$&$	0 	$&$	-	$&$	26.53 	$&$	26.47 	$&$	26.58 	$&$	26.36 	$&$	27.89 	$\\
			
			\hline
			\hline
	\end{longtable*}
\endgroup

\begin{figure}[h]\centering
	\includegraphics[width=9cm]{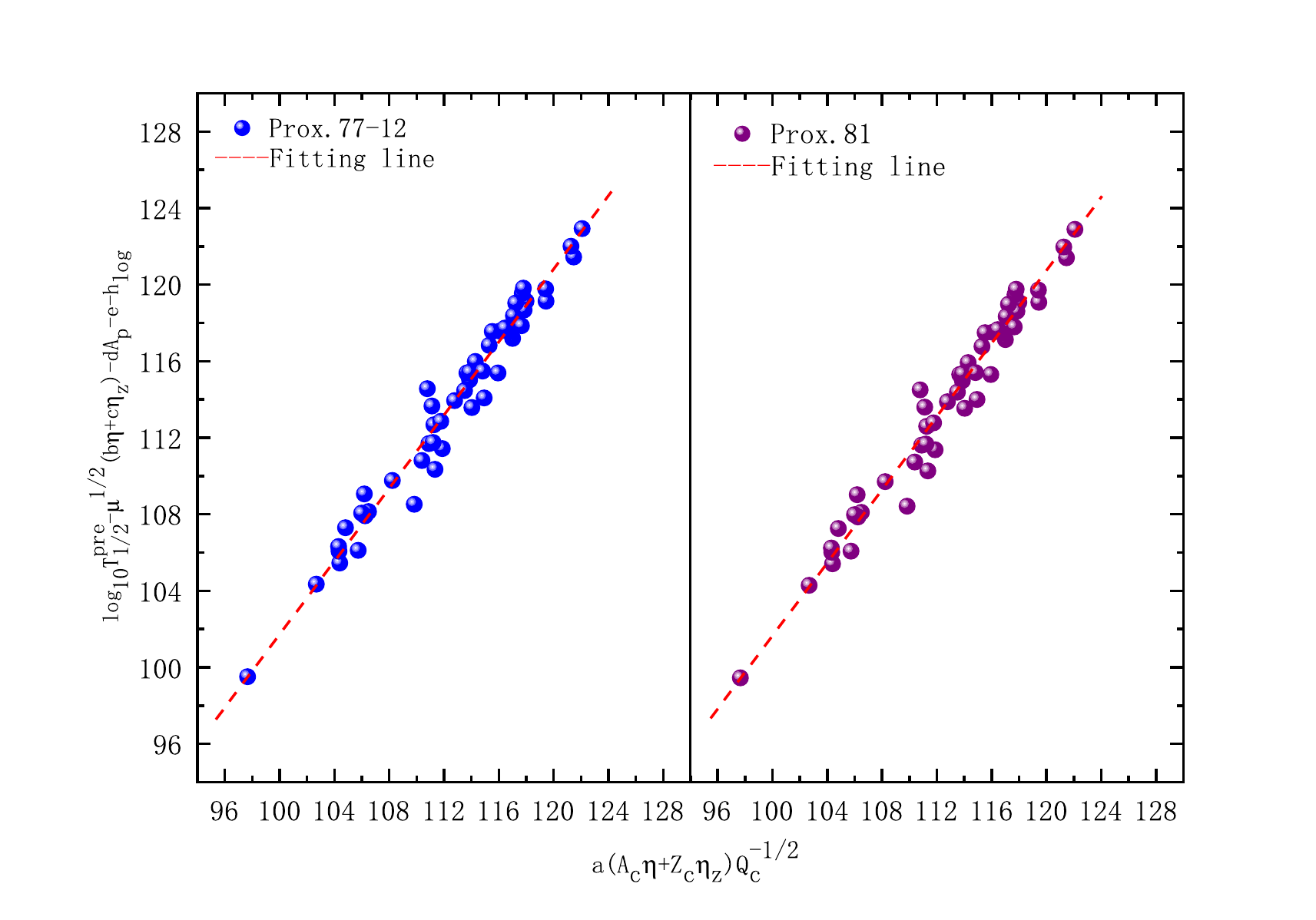}
	\caption{(color online) The linear relationship between the quantity $\log_{10}T_{1/2}^{pre}-{\mu}^{1/2}(b\eta+c\eta_{z})-dA_{p}-e-h_{log}$ and $a(A_{c}\eta+Z_{c}\eta_{z}){{Q_{c}}^{-1/2}}$ based on the empirical formula in Ref. \cite{2023CPC47064107}.}
	\label {fig 4}
\end{figure} 
Recent research has shown that the New Geiger-Nuttall law can be used to describe the
all cluster radioactivity within an empirical formula \cite{2023CPC47064107}, which is not just limited to isotopes. To further confirm the feasibility of our predictions, according to the formula in Ref. \cite{2023CPC47064107}, we plot the quantity $\log_{10}T_{1/2}^{pre}-{\mu}^{1/2}(b\eta+c\eta_{z})-dA_{p}-e-h_{log}$ as a function of $a(A_{c}\eta+Z_{c}\eta_{z}){{Q_{c}}^{-1/2}}$ in Fig.\ref{fig 4}. From this figure, we can find there is an obvious linear dependence of $\log_{10}T_{1/2}^{pre}$ on $Q_{c}$$^{-1/2}$ for Prox.77-12 and Prox.81 when other variables are treated as constants. It is demonstrated that our predictions are credible.

\section{Summary}
\label{sec:Summary}

In summary, a systematic comparative study was performed on 28 versions
of the proximity potential substituting the potential nuclear part to calculate the cluster radioactivity half-lives of 26 nuclei. The theoretical results were compared with the experimental data using the root-mean-square deviation, which is found that the proximity potential Prox.77-12 and Prox.81 formalisms give the lowest rms deviation $\sigma=0.681$ in the description of the experimental half-lives of the known cluster emitters. Furthermore, we use the CPPM with four proximity potential formalisms of the smallest $\sigma$
to predict the half-lives of 51 possible cluster radioactive candidates. These predicted results are in reasonable agreement with the calculated ones by using Ni's empirical formula. By the branching ratio $\varsigma$, it is found that the former to be more dominant in the competition between $\alpha$ decay and cluster radioactivity for these predicted nuclei. 
Meanwhile, we also use the new Geiger-Nuttall law to verify the viability of these predictions in cluster radioactivity. This work maybe provide an appropriate reference for experimental and theoretical research in the future.
 
\begin{acknowledgments}

This work is supported in part by the National Natural Science Foundation of China (Grant No.12175100 and No.11975132), the construct program of the key discipline in hunan province, the Research Foundation of Education Bureau of Hunan Province, China (Grant No.18A237 and No.22A0305), Hunan Provincial Department of Education Scientific Research Project (Grant No.19A440), the Shandong Province Natural Science Foundation, China (Grant No.ZR2022JQ04), the Opening Project of Cooperative Innovation Center for Nuclear Fuel Cycle Technology and Equipment, University of South China (Grant No.2019KFZ10), the Innovation Group of Nuclear and Particle Physics in USC, Hunan Provincial Innovation Foundation for Postgraduate (Grant No.CX20230962). Science and technology plan project of hengyang city (Grant No.202150063428).
\end{acknowledgments}

%\bibliographystyle{apsrev4-1}
%\bibliography{1st}

%merlin.mbs apsrev4-1.bst 2010-07-25 4.21a (PWD, AO, DPC) hacked
%Control: key (0)
%Control: author (72) initials jnrlst
%Control: editor formatted (1) identically to author
%Control: production of article title (-1) disabled
%Control: page (0) single
%Control: year (1) truncated
%Control: production of eprint (0) enabled
%

\end{document}